\documentclass[journal=jacsat,manuscript=article,layout=twocolumn]{achemso}
\usepackage[version=3]{mhchem}

\usepackage[T1]{fontenc} 
\usepackage{xcolor}
\usepackage{amsmath}
\usepackage{lipsum}
\usepackage{widetext}
\pdfoutput=1
\setkeys{acs}{maxauthors = 0}
\let\oldmaketitle\maketitle
\let\maketitle\relax

\usepackage{siunitx}
\DeclareSIUnit{\kT}{k_{B}T}
\DeclareSIUnit{\Molar}{\textsc{m}} 

\title{Quantitative Prediction of Protein--Polyelectrolyte Binding Thermodynamics: Adsorption of Heparin-Analog Polysulfates to the SARS-CoV-2 Spike Protein RBD}

\author{Lenard Neander}
\affiliation{Department of Physics, Freie Universität Berlin, Arnimallee 14, 14195 Berlin, Germany}
\alsoaffiliation{Institute of Chemistry and Biochemistry, Freie Universität Berlin, Takustraße 3, 14195 Berlin, Germany}
\author{Cedric Hannemann}
\affiliation{Department of Physics, Freie Universität Berlin, Arnimallee 14, 14195 Berlin, Germany}
\author{Roland R. Netz}
\affiliation{Department of Physics, Freie Universität Berlin, Arnimallee 14, 14195 Berlin, Germany}
\email{rnetz@physik.fu-berlin.de}
\author{Anil Kumar Sahoo}
\affiliation{Department of Physics, Freie Universität Berlin, Arnimallee 14, 14195 Berlin, Germany}
\email{aksahoo@zedat.fu-berlin.de}

\begin{document}


\twocolumn[
\begin{@twocolumnfalse}
\oldmaketitle
\begin{abstract}
Interactions of polyelectrolytes (PEs) with proteins play a crucial role in numerous biological processes, such as the internalization of virus particles into host cells. Although docking, machine learning methods, and molecular dynamics (MD) simulations are utilized to estimate binding poses and binding free energies of small-molecule drugs to proteins, quantitative prediction of the binding thermodynamics of PE-based drugs presents a significant obstacle in computer-aided drug design. This is due to the sluggish dynamics of PEs caused by their size and strong charge--charge correlations. In this paper, we introduce advanced sampling methods based on a force-spectroscopy set up and theoretical modeling to overcome this barrier. We exemplify our method with explicit solvent all-atom MD simulations of interactions of anionic PEs that show antiviral properties, namely heparin and linear polyglycerol sulfate (LPGS), with the SARS-CoV-2 spike protein receptor binding domain (RBD).  
Our prediction for the binding free-energy of LPGS to the wild-type RBD matches experimentally measured dissociation constants within thermal energy, $k_\mathrm{B}T$, and correctly reproduces the experimental PE-length dependence. 
We find that LPGS binds to the Delta-variant RBD with an additional free-energy gain of 2.4 $k_\mathrm{B}T$, compared to the wild-type RBD, in accord with electrostatic arguments.
We show that the LPGS--RBD binding is solvent dominated and enthalpy driven, though with a large entropy--enthalpy compensation. Our method is applicable to general polymer adsorption phenomena and predicts precise binding free energies and reconfigurational friction as needed for drug and drug-delivery design. 
\end{abstract}
\end{@twocolumnfalse}
]

\section{Introduction}
 Understanding the interaction of PEs with proteins is important for interpreting structure formation in biology, e.g., in nucleosomes \cite{kornberg1999twenty, kunze2002complexes, parsaeian2013binding}, intracellular condensates \cite{shin2017liquid}, and the brain,\cite{goedert1996assembly, fichou2018cofactors} as well as elucidating different biological processes including viral infections or inhibitions \cite{achazi2021understanding, kayitmazer2013protein, nandy2015spl7013}.
 The cell entry of many viruses is mediated by the interaction of anionic heparan sulfate proteoglycans (HSPGs) present on the extracellular matrix of the host cell with positively charged viral surface glycoproteins \cite{cagno2019heparan}. Due to this non-specific electrostatic attraction, the concentration of virions at the cell surface is increased, making it more likely for them to bind to the host cell receptor proteins, which often includes multivalent interactions \cite{liese2018quantitative} and eventually leads to viral invasion \cite{lauster2023respiratory}. This mechanism has been observed for different viruses such as Hepatitis B and C, herpes simplex virus, etc.\cite{cagno2019heparan} and more recently for the SARS-CoV-2 virus, responsible for the COVID-19 pandemic \cite{chu2021host}. In the case of SARS-CoV-2, a cationic patch present on the spike protein RBD binds to HSPGs \cite{Main_article}, which makes it more feasible for the RBD to specifically interact with the host cell receptor, the angiotensin-converting enzyme 2 (ACE2) \cite{Clausen_article}. \par{}
The study of PE--protein interaction is also of major importance for the development of new drugs. Heparin, a naturally occurring anionic PE, has been the subject of intense research in the last few decades due to its ability to adsorb at positively charged surfaces and thereby modulate biological processes \cite{heparin_non_coagulant, page2013heparin, heparin}. 
More recently it has been used to treat patients with SARS CoV-2 infections \cite{Bleeding_article, lan2020structure, Main_article, Clausen_article}. By binding to the cationic patch of the RBD, heparin can compete with HSPGs and block the first step of the cell-entry process. However, heparin-based drugs come with anticoagulatory side effects, which can be a disadvantage for the treatment of specific diseases \cite{pishko2019risk, Bleeding_article}. Therefore,  there is a great interest in developing heparin analogs that share the same characteristic for adsorbing to cationic surfaces but with less side effects. LPGS has recently been tested to show excellent inhibitory activity against SARS-CoV-2 \cite{Main_article}. Compared to heparin, LPGS shows larger binding affinities to the RBD of wild-type SARS-CoV-2 and its different variants \cite{Main_article, nie2022charge}.
\par{} 
Computational methods developed in the last few decades have been quite successful in predicting binding poses and binding free-energies of small-molecule drugs with proteins \cite{wang2015accurate, cournia2017relative, mobley2017predicting, zhao2020exploring, gapsys2021accurate}. 
However, quantitative prediction of polymer--protein binding thermodynamics remains challenging because of the limited lengthscales and timescales accessible by atomistic simulations using present-day computational power.\cite{xu2018interaction} Specifically, PE-based viral inhibitors tested in experiments typically are 100--1000mers long with molecular weights of 10--100 kDa \cite{Main_article, bhatia2017linear} and the binding--unbinding equilibrium relaxation time for interactions between a charged monomeric unit and an oppositely charged protein residue can be a few microseconds.
\par{}
In this article, we combine advanced sampling techniques and theoretical modeling to investigate the interaction of LPGS and heparin with the SARS-CoV-2 spike protein RBD using explicit solvent all-atom MD simulations. 
Adapting a 
simulation set-up that mimics atomic force microscopy experiments,\cite{Method_article, schwierz2012relationship} we determine the PE--protein adsorption free energy from measuring the force to pull away from the protein surface an adsorbed short (a few monomers-long) PE. The adsorption free-energy thus obtained is compared with that computed using umbrella sampling simulations for validation. To obtain the standard binding free-energy of PEs, we add two correction terms: 
the PE stretching free energy due to the applied force and an entropic term accounting for its binding volume. For comparing the simulation results with experiments where longer PEs have been considered, 
we add the polymer translational entropy contribution in the bound state,
which scales logarithmically in the degree of polymerization $N$, and find good agreement for the interaction of LPGS with the wild-type RBD compared with recent experimental measurements \cite{Main_article, page2023functionalized}. Moreover, we decompose the free energy into enthalpic and entropic contributions arising from solute--solute and solute--solvent interactions, for a deeper understanding of the PE--protein binding thermodynamics. \par{}
\section{Results and Discussion}
\begin{figure*}[h!]
    \centering
    \includegraphics[width=.95\textwidth]{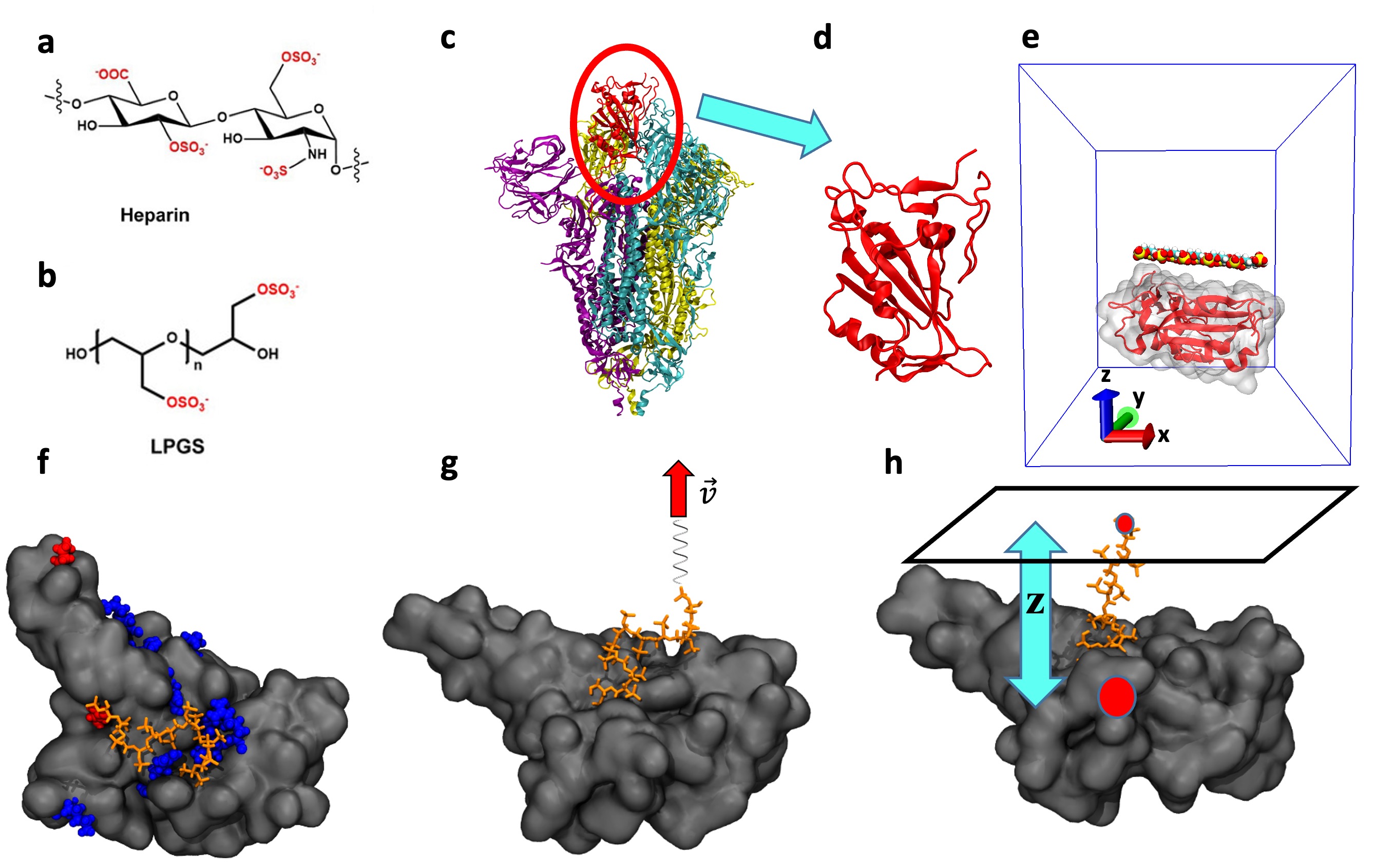}
     \caption{Simulation details. Chemical structures of \textbf{(a)} heparin and \textbf{(b)} LPGS. \textbf{(c)} Structure of the spike protein trimer in the secondary structure representation (PDB ID: 7DK3)\cite{xu2021conformational}. Monomers are shown in purple, cyan, and yellow. The receptor binding domain (RBD) of one monomer (cyan) is present in the up conformation and is shown in red. \textbf{(d)} Zoomed-in view of the RBD. \textbf{(e)} Simulation unit cell (blue box) for equilibrium adsorption of a polymer to the RBD (red).
     An LPGS undecamer is shown in the space-filling representation with color coding for the different atom types: hydrogen in white, carbon in cyan, oxygen in red, and sulfur in yellow. Water and ions are present but not shown for clarity. The x, y, and z axes are indicated with red, green, and blue arrows, respectively. \textbf{(f)} Simulation snapshot after \SI{1000}{\nano\second} of equilibration representing adsorption of the LPGS (shown in orange) to the Delta-variant RBD surface (shown in grey). Protein cationic residues are pointed out in blue, whereas mutated residues R452 and L478 are in red. \textbf{(g)} Dynamic pulling simulation protocol in which a spring connected to one of the terminal atoms of the polymer is pulled away from the protein surface with a constant velocity $v$ along the z-axis. The spring is free to move along the lateral direction. \textbf{(h)} Static pulling simulation protocol in which one of the terminal atoms of the polymer is allowed to freely move in a plane at constant z-separation (distance projected along the normal to the plane) from the protein center-of-mass (red circle).}
    \label{fig:intro_snaps}
\end{figure*}

Due to the large size of the trimeric SARS-CoV-2 spike protein, we consider in simulations only one monomer's 
RBD (Figure~\ref{fig:intro_snaps}c,d). The RBD binds to not only the HSPGs on the 
cell surfaces but also to the cell receptor protein ACE2. Thus, it plays a central role in the cell entry process of the 
virion and constitutes a suitable target for new drugs. Moreover, dissociation constants for PEs binding to RBD have 
been reported from microscale thermophoresis experiments \cite{Main_article, page2023functionalized}, 
making a meaningful comparison with the simulations possible. It should be noted that in vivo, there are
glycans attached to the RBD \cite{woo2020developing, casalino2020beyond}, which are not considered in our
simulations as these are present on the opposite side of the RBD surface containing the cationic patch, 
the putative sulfate binding sites \cite{Main_article, nie2022charge}, and hence should not affect the binding of anionic PEs to the RBD. 
We perform simulations of LPGS (see the chemical structure in Figure~\ref{fig:intro_snaps}b) interacting with both the 
wild-type RBD and the Delta-variant RBD with L452R and T478K mutations. We take an LPGS undecamer in the simulations, much shorter than that used in experiments, to ensure fast equilibration. Additionally, we 
conduct simulations of a heparin pentamer (chemical structure of the monomer shown in Figure~\ref{fig:intro_snaps}a) 
interaction with the wild-type RBD, for which we observe a very long equilibration time, 
as will be discussed in Figure~\ref{fig:SP}b,d. Therefore, in the analysis we focus on simulation results 
involving LPGS binding and their quantitative comparison with experiments. \par{}
We start with unrestrained simulations (the set-up shown in Figure~\ref{fig:intro_snaps}e), where a polymer can move freely in the simulation box and interact with the protein surface. We observe that LPGS binds to the cationic patch of both the wild-type and Delta-variant RBD (shown in Figure~\ref{fig:intro_snaps}f in blue). To validate that the adsorption to the cationic patch corresponds to the optimal binding position and is not an artifact of the starting position, we conduct five independent simulations, where we place LPGS at different starting positions. We find that within \SI{500}{\nano\second} LPGS binds to the same cationic patch in all five simulations, 
as shown in Figure S4 in the supporting information (SI). Therefore, the observed binding conformation is expected to be the minimum free energy configuration and is suitable as a starting structure for subsequent pulling simulations.\par{}
\subsection{Polymer Desorption Free Energy}
\begin{figure*}[h]
    \centering
    \includegraphics[width=0.95\textwidth]{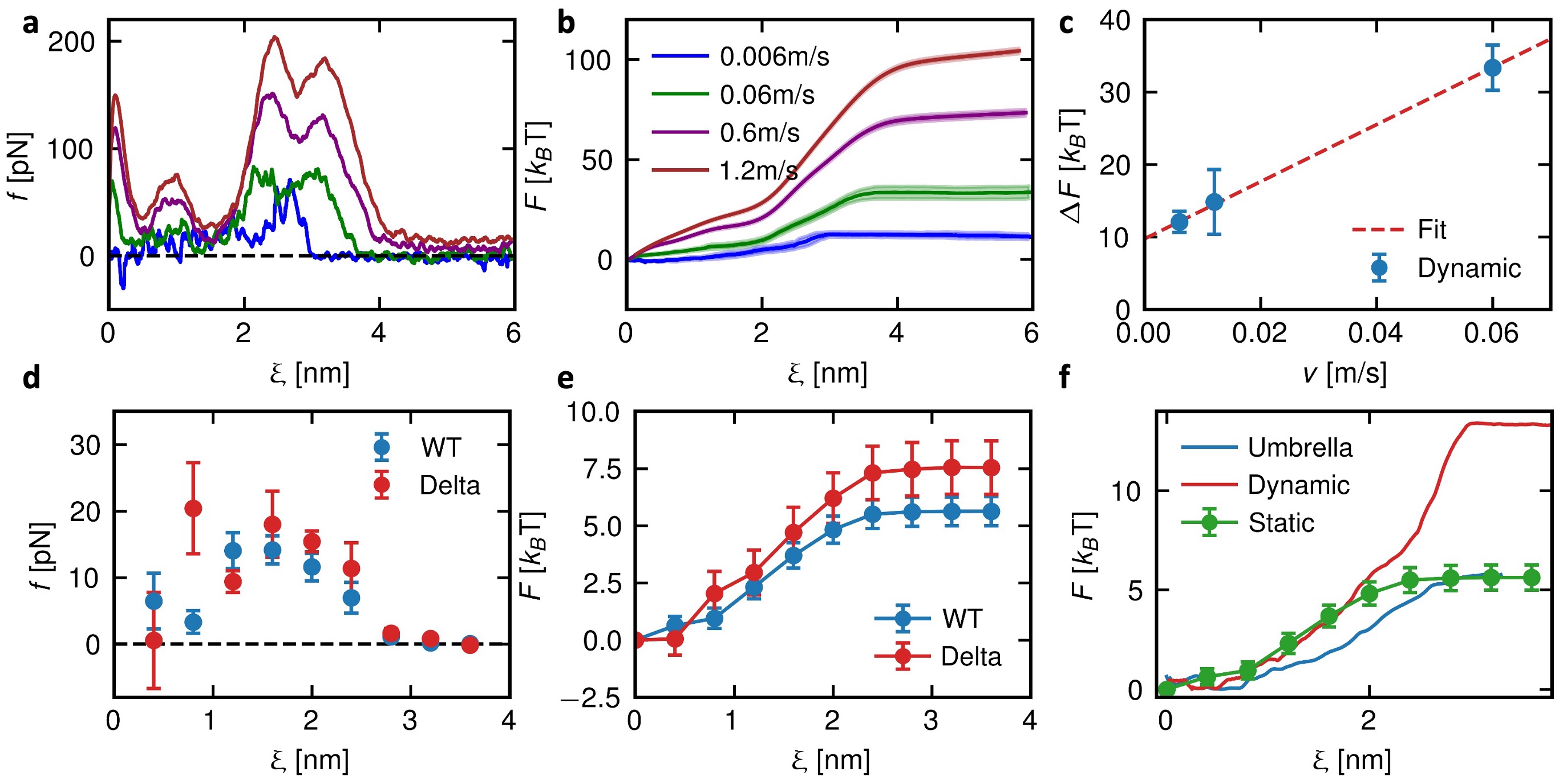}
    \caption{Simulation results for the desorption of an LPGS undecamer from the wild-type (WT) and, only if mentioned, Delta-variant RBD surface. \textbf{(a)} Force $f$ and \textbf{(b)} free energy $F$ profiles from dynamic pulling simulations, results for only four out of the six pulling velocities are displayed for visual clarity. Error bars are displayed as shaded colored areas in panel \textbf{b}. \textbf{(c)} Free-energy difference $\Delta F$ between the desorbed and adsorbed states as a function of the pulling velocity $v$. The equilibrium free-energy difference $\Delta F (v=0)$ is obtained from linear extrapolation of the data for the three slowest pulling rates to zero velocity. \textbf{(d)} Force and \textbf{(e)} free-energy profiles from static pulling simulations. \textbf{(f)} Comparison of free energy profiles obtained from umbrella sampling and static pulling simulations with the dynamic pulling simulation result using the slowest pulling velocity $v =$ \SI{0.006}{\meter\per\second}.
    }
    \label{fig:simulation_results}
\end{figure*}

To obtain the LPGS desorption free energy profile as a function of the reaction coordinate $\xi$, defined as
the 
pulled distance of the polymer terminus from the protein surface
(see Figure~\ref{fig:intro_snaps}h), we use three different sampling techniques: dynamic pulling, static pulling, and umbrella sampling.
In the dynamic pulling simulations, a harmonic spring attached to one of the polymer's terminal atoms is moved away from the protein surface along $\xi$ at speed $v$ (see Figure~\ref{fig:intro_snaps}g) and the force $f$ acting on the spring is measured. Starting with the system configuration of LPGS adsorbed to the wild-type RBD, simulations are conducted with six different speeds 
ranging $v =$ \SI{0.006}{\meter\per\second} to \SI{1.2}{\meter\per\second}. To make sampling comparable for different $v$, we perform multiple simulations such that a total simulation time of at least \SI{1}{\micro\second} is reached for each $v$. 
Desorption force profiles $f(\xi)$ for different $v$ are shown in Figure~\ref{fig:simulation_results}a. We find that the most distinct force peaks are situated at $\xi$ in the range of 2--\SI{4}{\nano\meter} and the desorption forces are higher for faster pulling speeds. The latter observation can be rationalized by friction effects originating from the breaking of hydrogen bonds and salt bridges \cite{Friction_article_1, Friction_article_2, patil2014viscous}. 
Free energy profiles $F(\xi)$, obtained by integrating over the force profile 
as
\begin{equation}
\label{formula:Path_integral_f}
    F(\xi) = \int \limits_{0}^{\xi} f(\xi') d\xi' , 
\end{equation} 
are shown for different pulling speeds in Figure~\ref{fig:simulation_results}b. 
We find that apart from the higher free energy values for faster $v$, the profiles for higher $v$ ($=$ \SI{1.2}{\meter\per\second} and \SI{0.6}{\meter\per\second}) do not reach plateau values, even after the complete desorption of the polymer. This shows that simulations are far from equilibrium for higher $v$. As the friction contribution to the desorption force $f$ in the viscous (i.e. low-velocity) regime is linear in $v$, the free-energy difference between the desorbed and adsorbed state, $\Delta F = F(\xi \to \infty)$, increases linearly with $v$ according to $\Delta F(v) = \Delta F(v=0) + \gamma v L_0$, where $L_0 = 4.03$ nm (see Methods) is the unstretched contour length of the polymer \cite{Friction_article_1}. 
We use this linear relationship to determine the equilibrium free-energy of polymer desorption, $\Delta F(v=0)$, and the friction coefficient $\gamma$. 
Linear regression, using $\Delta F$ data for only the three lowest $v$, leads to an excellent fit (Figure \ref{fig:simulation_results}c), resulting in $\Delta F(v=0) = 9.7 \pm 1.6\ k_\mathrm{B}T$ 
and $\gamma = 4.05 \times 10^{-10}$ Ns/m
(for results from fitting the whole dataset, see Figure S5 in the SI). 
This is equivalent to a diffusion constant of $D = k_\mathrm{B}T/\gamma = 10.23\ \mathrm{nm^2/{\mu s}}$, from which we obtain the diffusion time $\tau_\mathrm{D} = L_0^2/2D = 0.8\ \mathrm{\mu s}$. We see that the bound polymer diffuses over its contour length in about a microsecond, allowing for rather quick equilibration of the bound state. \par{}
For the static pulling simulations, we select initial configurations from the dynamic pulling simulation at nine different $\xi$ values with a spacing of \SI{0.4}{\nano\meter}. 
At each $\xi$ value, the polymer terminal atom is allowed only to freely move on a plane, maintaining a constant z-separation (distance projected along the normal to the plane) from the protein surface, see Figure~\ref{fig:intro_snaps}h. 
The average force needed to keep the LPGS terminal atom at different $\xi$ values is shown in Figure~\ref{fig:simulation_results}d for the wild-type and Delta-variant RBD. 
Compared to dynamic pulling, the peak force here is significantly lower and the force drops to zero already at smaller $\xi$ values, signifying the complete desorption of the polymer. These deviations illustrate that even for the slowest pulling rate, dynamic simulations are far from equilibrium. 
Moreover, it is observed that forces for the Delta variant are significantly higher than the wild-type RBD at $\xi=\SI{0.8}{\nano\meter}$. To understand this, we investigate the interaction of LPGS with individual protein residues 
(see Figure 
S6 in the SI). For both RBD types, we find that LPGS forms a higher number of contacts with cationic residues such as R346 and R466. For the Delta variant, there is however an extra charged residue on the cationic patch (R452), which makes additional, electrostatically favorable contacts with LPGS simultaneously possible 
(Figure S6d in the SI), giving rise to the increased force.
Free energy profiles 
show that the complete desorption of LPGS from the Delta-variant RBD, compared to the wild-type, requires an extra free energy of \SI{2}{\kT} (Figure~\ref{fig:simulation_results}e). This demonstrates that cationic mutations in proteins lead to an increase in their binding affinities to anionic polymers, which supports the hypothesis of Nie et al.\ \cite{nie2022charge} that the increased infectivities of Delta and Omicron variants are caused by additional cationic residues of the RBD that interacts strongly with cellular HSPGs. \par{}
We also calculate the free energy profile of LPGS desorption from the wild-type RBD, using umbrella sampling simulations with the weighted histogram analysis method (for details, see Methods). \cite{kumar1992weighted, torrie1977nonphysical} Free energy profiles $F(\xi)$ obtained from all three methods are compared in Figure~\ref{fig:simulation_results}f. 
Although $F(\xi)$ from the dynamic pulling simulation using even the slowest pulling velocity, $v=\SI{0.006}{m/s}$, deviates significantly from the other two methods, the extrapolated free-energy difference $\Delta F (v=0)$ to zero velocity agrees better with the umbrella sampling, $\Delta F = 5.8 \pm 0.5\ k_\mathrm{B}T$, and static pulling, $\Delta F = 5.6 \pm 0.6\ k_\mathrm{B}T$, results.
\par{}
\subsection{Dissociation Constant and Standard Binding Free Energy}
The dissociation constant $K_\mathrm{D}$ for a complexation reaction between protein $P$ and polymeric ligand $L$  
\begin{equation}
    P + L \rightleftharpoons PL
\end{equation}
determines the ratio of concentrations of free, $[P]$, $[L]$, and bound, $[PL]$, species in a solution and is related to the free-energy change upon binding, $\Delta F_\mathrm{b}$, and the ligand binding volume $V_\mathrm{b}$ according to \cite{phillips2012physical}
\begin{equation}  \label{formula:kd}
    K_\mathrm{D} = \frac{[P][L]}{[PL]}= \frac{1}{V_\mathrm{b}}\mathrm{e}^{\Delta F_\mathrm{b}/ k_\mathrm{B}T} = \frac{1}{V_\mathrm{0}} \mathrm{e}^{\Delta F_\mathrm{b}^0 / k_\mathrm{B}T}.
\end{equation}
Here, $V_\mathrm{b}$ represents the limited volume available for the polymer to move in the protein-bound state (for the procedure to obtain $V_\mathrm{b}$ from simulations, see the SI text and Figs. S2 and S3).
$V_0 = 1.661\ \mathrm{nm}^3$ is the standard-state volume corresponding to the concentration of 1 M and the standard free-energy of binding is given by $\Delta F_\mathrm{b}^0 = \Delta F_\mathrm{b} -  k_\mathrm{B}T\ln\left(V_\mathrm{b}\right / V_0)$.
$\Delta F_\mathrm{b}$ is 
obtained from the polymer desorption free-energy $\Delta F$ 
by removing the polymer stretching free-energy due to the applied force, resulting in 
\begin{equation}
\Delta F_{\mathrm{b}}= -\Delta F - \Delta F_{\mathrm{stretch}} . 
\end{equation}
\par
\begin{figure}[h!]
    \centering
    \includegraphics[width=0.445\textwidth]{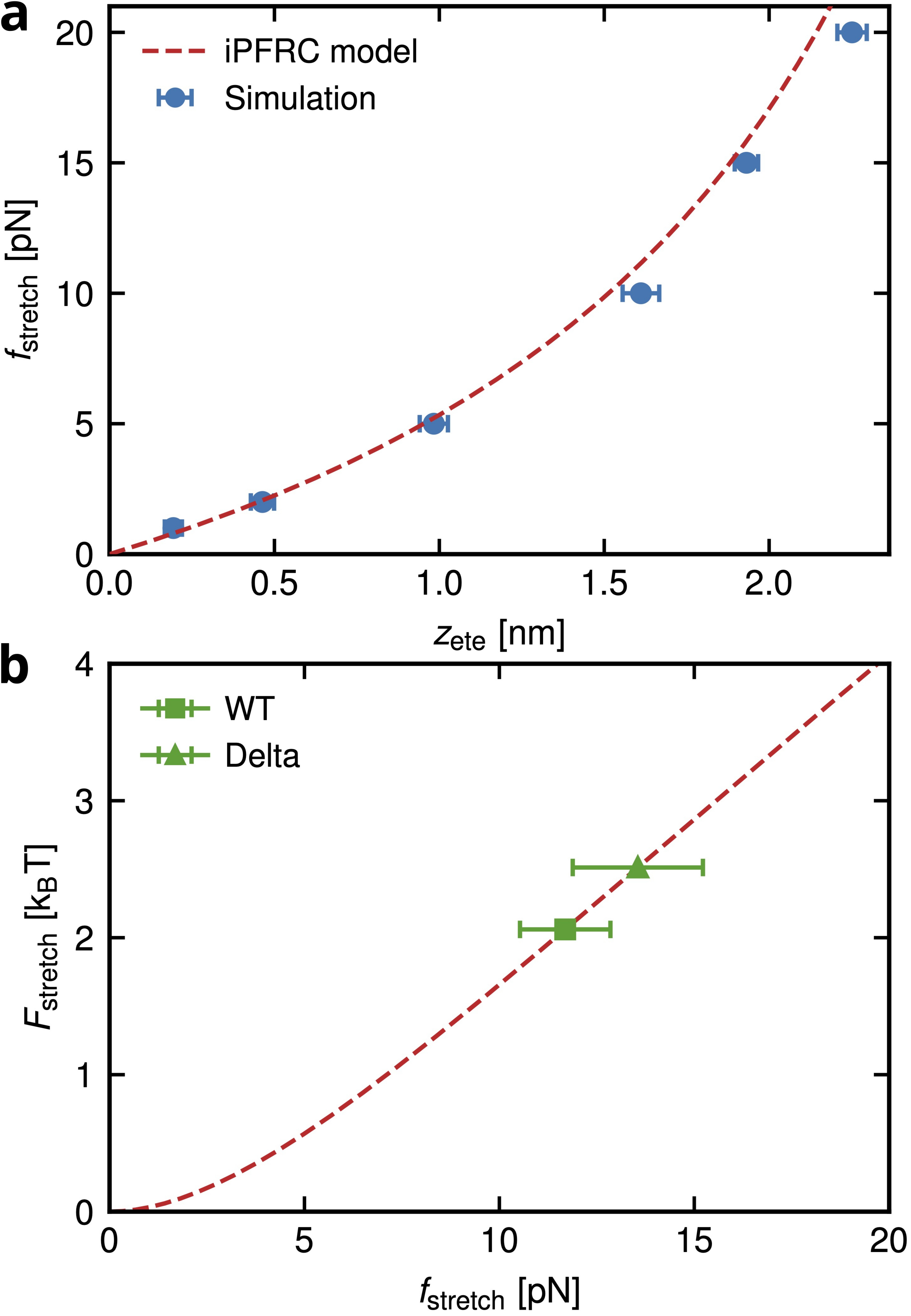}
    \caption{\textbf{(a)} Stretching force $f_{\mathrm{stretch}}$ versus the average end-to-end distance $z_{\mathrm{ete}}$ of an LPGS undecamer along the direction of applied force. Data points are fitted to the iPFRC model force--extension relation Eq. \ref{eq:f-e_iPFRC}. \textbf{(b)} Stretching free-energy profile $F_{\mathrm{stretch}}$ 
    as a function of $f_\mathrm{stretch}$, obtained by integrating the fitted curve in panel \textbf{a}.  
    }
    \label{fig:stretching_free_Energy}
\end{figure}

{}
$\Delta F_{\mathrm{stretch}}$ describes the change in free energy to
stretch the polymer from its relaxed state and 
is not present in the experiments and hence its contribution is subtracted from the simulated free energy.
%
by measuring the average end-to-end distances, $z_{\mathrm{ete}}$, projected along the direction of external force at different stretching forces, $f_{\mathrm{stretch}}$, applied to each end of the polymer. We fit the force--extension relation of the inhomogeneous partially freely rotating chain (iPFRC) model \cite{hankeStretchingSinglePolypeptides2010} (for details, see Methods) to the simulation values and find an excellent fitting as shown in Figure~\ref{fig:stretching_free_Energy}a.
By integrating the force--extension relation from the relaxed end-to-end distance $z_{\mathrm{ete}}^0 = 0$ to a stretched end-to-end distance $z_{\mathrm{ete}}^{\mathrm{stretch}}$, 
\begin{equation}
\Delta F_{\mathrm{stretch}} = {\int \limits_{z_{\mathrm{ete}}^{0}}^{z_{\mathrm{ete}}^{\mathrm{stretch}}} f_{\mathrm{stretch}}(z_{\mathrm{ete}})dz_{\mathrm{ete}}},  \label{eq:u_stretch}
\end{equation} 
we obtain the stretching free energy profile as a function of $f_{\mathrm{stretch}}$, shown in Figure~\ref{fig:stretching_free_Energy}b. To obtain $\Delta F_\mathrm{stretch}$, we compute the average force experienced by a strained LPGS bound to the RBD (see the force plateau in Figure~\ref{fig:simulation_results}d from \SI{1.2}{\nano\meter} to \SI{2.4}{\nano\meter}) and take the stretching free energy at the average forces of $f_\mathrm{stretch} = 11.7$ pN and 13.5 pN for the wild-type and Delta RBD, respectively (see Figure~\ref{fig:stretching_free_Energy}b, green points).
Note that the polymer stretching free-energy contribution, $\Delta F_{\mathrm{stretch}} \simeq$ 2-2.5 $k_\mathrm{B}T$, is sizeable.
\par{} 
Finally, the standard binding free-energy $\Delta F_\mathrm{b}^0$ as a function of the degree of polymerization, $N$, for longer polymers used in the experiments  \cite{Main_article, page2023functionalized}, compared to the simulations, $N > N_\mathrm{sim}$, can be extrapolated from simulations data as (for a derivation, see the SI text)
\begin{widetext}
\begin{align}\label{eq:f_bind_long}
    \Delta F_{\mathrm{b}}^{0}(N) = -\Delta F(N_\mathrm{sim}) -\Delta F_{\mathrm{stretch}}(N_\mathrm{sim}) -k_{B}T\ln \left( \frac{V_\mathrm{b}(N_\mathrm{sim})}{V_0} \right)  -k_{B}T\ln \left( \frac{N}{N_\mathrm{sim}} \right) .
\end{align}
\end{widetext}
Here, the last term represents the \textit{avidity entropy} contribution, $-T\Delta S_\mathrm{avidity}$, to the binding, as explained below. \cite{kitov2003nature, bhatia2017linear, zumbro2019computational}
The direct polymer--protein interaction energy contribution to the binding free energy is limited to the number of binding sites $n_\mathrm{b} \simeq 5$ on the protein, estimated to be the number of charged residues on the cationic patch of the RBD. Thus, LPGS longer than a critical length, equivalent to the size of the cationic patch, do not contribute to this direct interaction. There is, however, a combinatorial entropy contribution for a longer polymer with $N > n_\mathrm{b}$, because of the different ways the polymer can bind to the protein, given by 
$S_\mathrm{avidity} = k_B\ln(\Omega_\mathrm{avidity})= k_B\ln (N-n_b+1)$. 
As this entropy contribution, $S_\mathrm{avidity}^\mathrm{sim} = k_B\ln (N_\mathrm{sim}-n_b+1)$, is already accounted for in our simulations, we obtain in the limit of $N > N_\mathrm{sim} >> n_\mathrm{b}$, $\Delta S_\mathrm{avidity} = k_B\ln(N/N_\mathrm{sim})$. \par{}
For the binding of the simulated LPGS undecamer to the wild-type and Delta-variant RBD, values of $\Delta F^{0}_\mathrm{b} (N_\mathrm{sim}) = -\Delta F -\Delta F_{\mathrm{stretch}} -k_{B}T\ln (V_\mathrm{b}/V_0)$, as the avidity entropy term in Eq. \ref{eq:f_bind_long} vanishes for $N = N_\mathrm{sim}$, and its different contributions are provided in Table~\ref{tab:params_simulation}. The LPGS undecamer binds to the Delta RBD more strongly than the wild-type with an additional free-energy gain of around $2.4\ k_\mathrm{B}T$, as expected from electrostatic arguments.
\par
\begin{figure*}[h!]
    \centering
    \includegraphics[width=.92\textwidth]{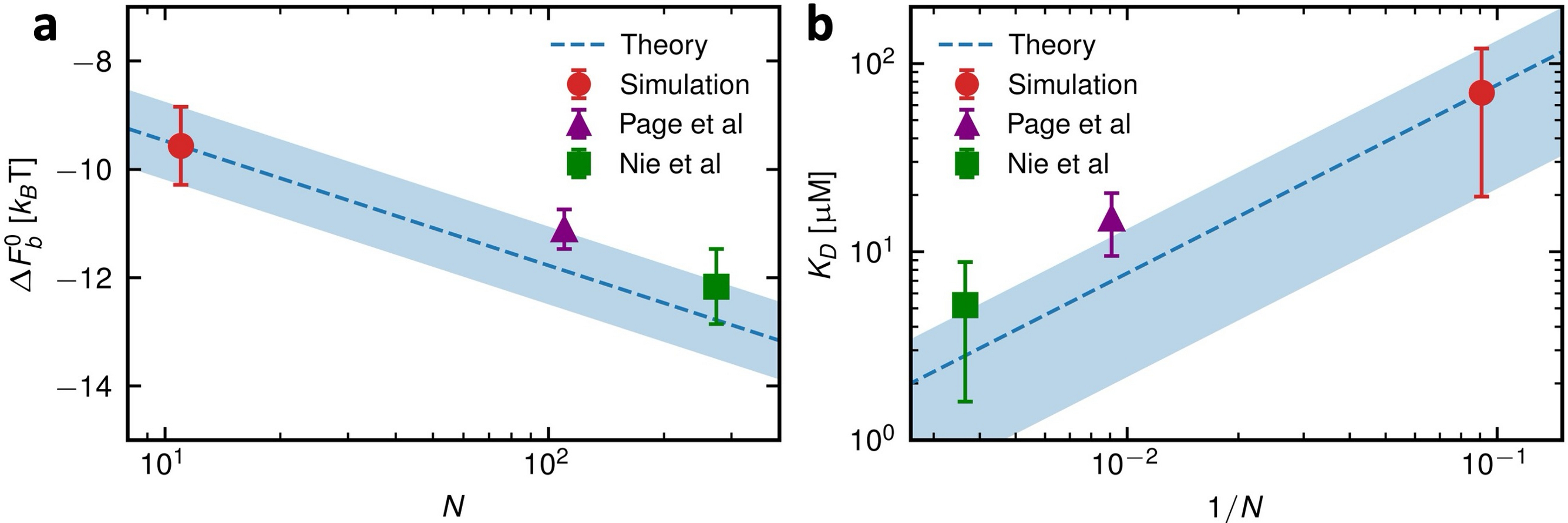}
    \caption{Theoretical prediction for \textbf{(a)} the standard binding free-energy $\Delta F_\mathrm{b}^0$ and \textbf{(b)} dissociation constant $K_\mathrm{D}$ for the complexation reaction of the wild-type RBD and LPGS as a function of its degree of polymerization $N$. Experimental values are taken from the publications by Nie et al.\ \cite{Main_article} and Page et al.\ \cite{page2023functionalized}. Shaded regions in panels \textbf{a} and \textbf{b} represent simulation errors of propagation coming from the first three terms in Eq. \ref{eq:f_bind_long} and from $\Delta F_\mathrm{b}^0$, respectively.
    }
    \label{fig:kd_extra}
\end{figure*}

{}
\begin{table*}[]
    \centering
    \begin{tabular}{|c|c|c|}
    \hline
     Contributions & WT & Delta \\
     \hline
       $\Delta F$ & \SI{5.59+-0.63}{\kT} & \SI{7.46+-1.17}{\kT}  \\
       \hline
        $\Delta F_{\mathrm{stretch}}$ & \SI{2.06+-0.29}{\kT} & \SI{2.51+-0.41}{\kT}\\
       \hline
        $V_{b}$ & \SI{11.32+-2.02}{\nano\meter^3}  & \SI{11.89+-4.91}{\nano\meter^3}\\
       \hline
       $k_\mathrm{B}T\ln\left(V_\mathrm{b}\right / V_0)$ & \SI{1.92+-0.11}{\kT}   & \SI{1.97+-0.25}{\kT} \\
       \hline
        $\Delta F_\mathrm{b}^0$ & \SI{-9.57+-0.72}{\kT} & \SI{-11.94+-1.30}{\kT}  \\
       \hline
    \end{tabular}
    \caption[Modelparameters]{Values for the standard free-energy of binding $\Delta F_\mathrm{b}^0$ of an LPGS undecamer, $N_\mathrm{sim}=11$, to the wild-type and Delta-variant RBD, along with the contributions in Eq.~\eqref{eq:f_bind_long} 
    to calculate $\Delta F_\mathrm{b}^0$.}
    \label{tab:params_simulation}
\end{table*}
The standard binding free energy $\Delta F_{\mathrm{b}}^{0}$ and dissociation constant $K_\mathrm{D}$ for different polymer lengths predicted according to Eq. \ref{eq:f_bind_long} and Eq. \ref{formula:kd} match nicely with 
the corresponding experimental values, as shown in Figure~\ref{fig:kd_extra} and Table \ref{tab:kd}. The theoretical and experimental $\Delta F_{\mathrm{b}}^{0}$ values differ only within 1 $k_\mathrm{B}T$ (the thermal energy), consequently $K_\mathrm{D}$ differ by around two times. \par{}
\begin{table*}[h!]
    \centering
    \begin{tabular}{|c|c|c|c|c|}
    \hline
       $N_\mathrm{exp}$ & $\Delta F_\mathrm{b}^0$ (exp.)\textsuperscript{\emph{a}} & $\Delta F_\mathrm{b}^0$ (theory)\textsuperscript{\emph{a}} &  $K_\mathrm{D}$ (exp.)\textsuperscript{\emph{b}} & $K_\mathrm{D}$ (theory)\textsuperscript{\emph{b}} \\
       \hline
       110 (Page et al.) & $-11.11\pm0.37$ & $-11.87\pm0.72$ & $15.0\pm5.5$ & $7.0\pm5.0$ \\
       \hline
       274 (Nie et al.) & $-12.17\pm0.69$ & $-12.78\pm0.72$ & $5.2\pm3.6$ & $2.8\pm2.0$ \\
       \hline
    \end{tabular}

    \textsuperscript{\emph{a}}Standard binding free-energy, $\Delta F^{0}_\mathrm{b}$ in $k_\mathrm{B}T$;
    \textsuperscript{\emph{b}}Dissociation constant, $K_\mathrm{D}$ in $\mathrm{\mu M}$
    \caption{Comparison of $\Delta F_\mathrm{b}^0$ and $K_\mathrm{D}$ values from experiments (Nie et al.\ \cite{Main_article} and Page et al.\ \cite{page2023functionalized}) and theoretical predictions for the wild-type RBD and LPGS binding for different experimental degrees of polymerization $N_\mathrm{exp}$.}
    \label{tab:kd}
\end{table*}
\subsection{Enthalpy--Entropy Decomposition}
\begin{figure*}[h!]
    \centering
    \includegraphics[width=.99\textwidth]{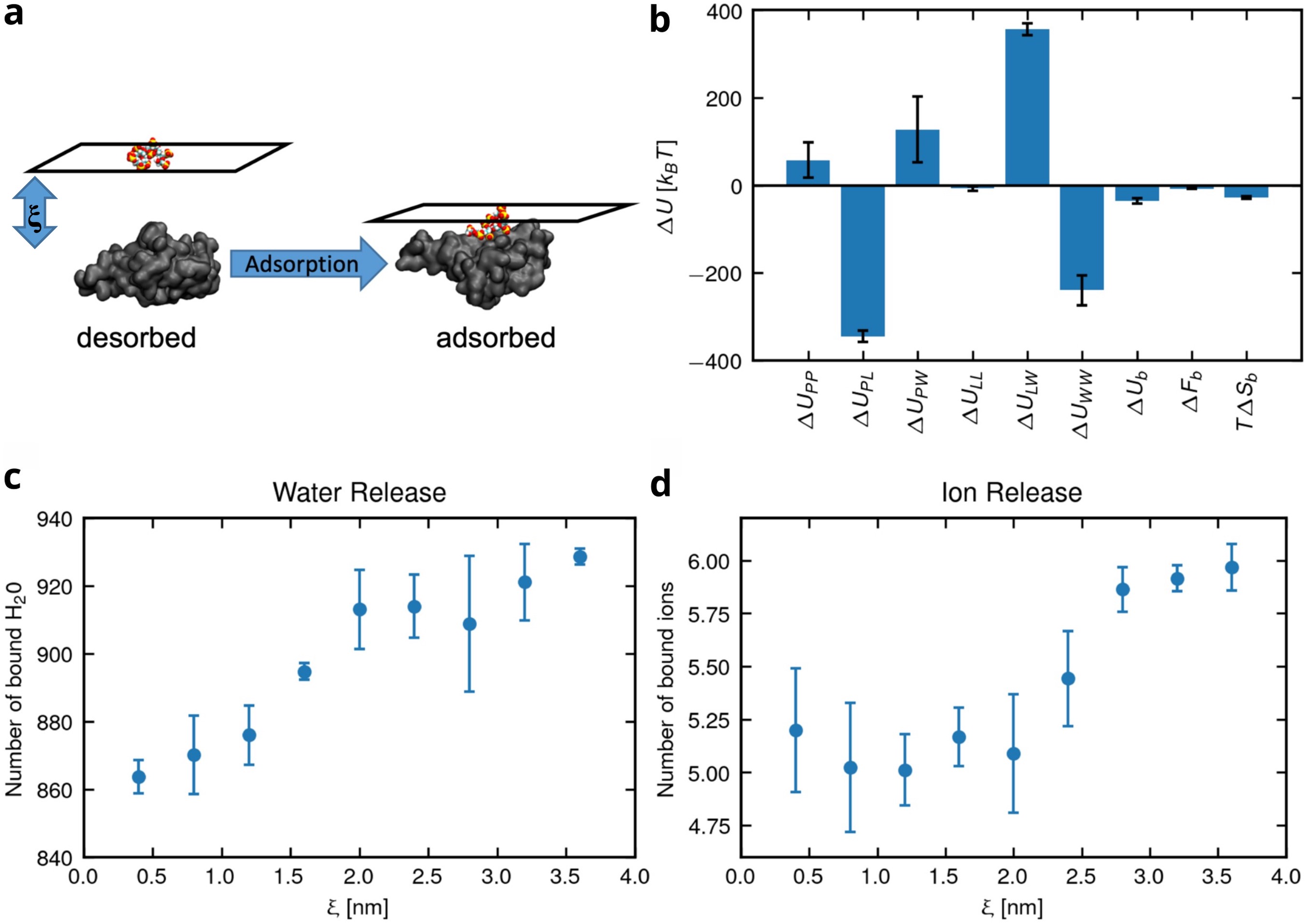}
    \caption{\textbf{(a)} Snapshots from the static pulling simulations for the desorbed state (at the pulled distance $\xi =$ \SI{3.6}{\nano\meter}) 
    and the adsorbed state ($\xi =$ \SI{0}{\nano\meter}).
    \textbf{(b)} Enthalpic, $\Delta U_\mathrm{b}$, and entropic, $T\Delta S_\mathrm{b}$, contributions to the binding free-energy $\Delta F_\mathrm{b}$ of the wild-type RBD--LPGS complex, along with the different internal energy contributions to $\Delta U_\mathrm{b}$ (see text).
    The average number of \textbf{(c)} water molecules and \textbf{(d)} ions, $\mathrm{Na^{+}}$ and $\mathrm{Cl^{-}}$, bound to the RBD or LPGS as a function of  $\xi$.}
    \label{fig:internal_energy}
\end{figure*}

To understand the underlying contributions to the binding free energy $\Delta F_\mathrm{b} = \Delta U_\mathrm{b} - T\Delta S_\mathrm{b}$, we decompose it into its enthalpic $\Delta U_\mathrm{b}$ and entropic $T\Delta S_\mathrm{b}$ parts, see Figure~\ref{fig:internal_energy}a,b. 
From the simulation trajectories (see Methods), we calculate the net change in interaction energy, $\Delta U_\mathrm{b} = \Delta U_{PP} + \Delta U_{PL} + \Delta U_{PW} + \Delta U_{LL} + \Delta U_{LW} + \Delta U_{WW}$, for the transition of the polymer from the desorbed to the adsorbed state from interactions among different components of the system: protein $P$, polymeric ligand $L$, and solvent (water molecules and ions) W.
In the polymer adsorption process, we find favorable intersolute direct interaction ($\Delta U_{PL} < 0$) and solvent reorganization energy ($\Delta U_{WW} < 0$) and unfavorable solute--solvent interactions ($\Delta U_{PW} > 0$ and  $\Delta U_{LW} > 0$). 
$\Delta U_\mathrm{b}$ is quite small compared to the different contributions. Thus, there are huge cancellations among the different interaction energy contributions, as seen for other receptor--ligand systems \cite{Method_article}, calling for highly accurate calculations of the solvent contribution \cite{qiao2019water}. \par{}
We obtain the entropy contribution to the binding using the thermodynamic relation $\Delta S_\mathrm{b} = (\Delta U_\mathrm{b} - \Delta F_\mathrm{b})/T$. By observing that $\Delta U_\mathrm{b} < T\Delta S_\mathrm{b} < 0$, we conclude that the adsorption process is entropically unfavorable and enthalpy driven. In the adsorption process, we find that 50--60 water molecules and only one ion are released as shown in Figure~\ref{fig:internal_energy}c,d (for the definition of bound water and ions, see Methods). Thus, water release is expected to contribute significantly to the entropy as gain, while the contribution due to counterion release is minimal. The latter is not surprising though, as the linear charge density of LPGS is just above the counterion condensation limit by Manning \cite{manning1969limiting}, the length of the simulated LPGS is short, thus leading to end effects \cite{fenley1990approach}, and the simulated salt concentration of 150 mM is high \cite{walkowiak2021interaction}. However, the net change in the binding entropy is negative (Figure~\ref{fig:internal_energy}b), which has to arise from the restricted conformational, translational, and rotational degrees of freedom of the polymer and protein in the bound state, overcompensating the entropy gain due to water and ion release.\cite{irudayam2009entropic, xu2018counterion, walkowiak2021interaction, caro2017entropy} 
As the release of a single water molecule from a typical protein surface leads to an entropy gain of $\sim\SI{1}{\kT}$ \cite{sahoo2022role}, the entropic loss due to conformational transformations of the protein and polymer is suggested to be greater than \SI{60}{\kT}.\par{}  
\subsection{Relaxation Time for Binding of Charged Groups} 
\begin{figure*}[h!]
    \centering
    \includegraphics[width=1.00\textwidth]{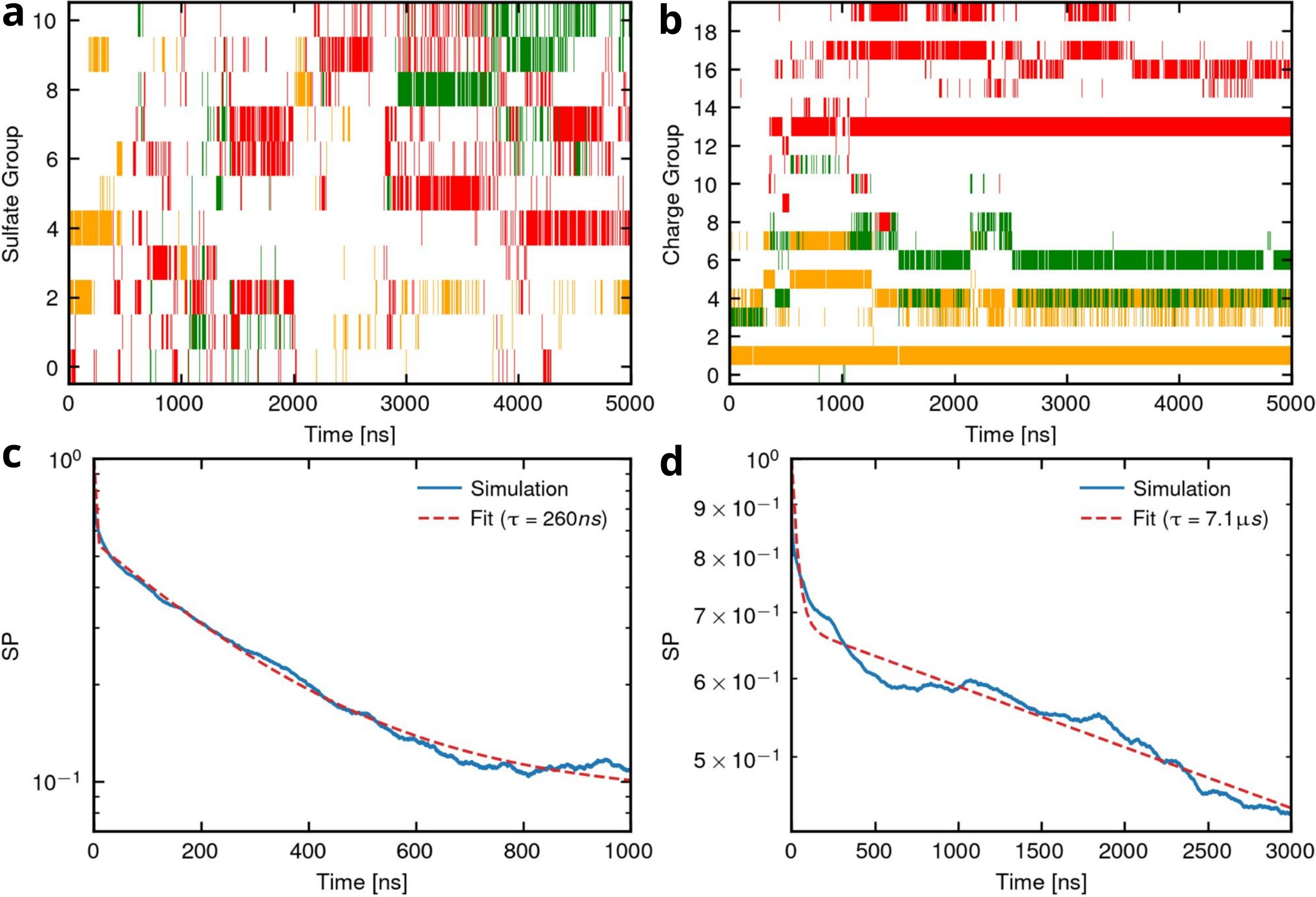}
    \caption{
    Time series for the binding of \textbf{(a)} LPGS's and \textbf{(b)} heparin's anionic groups (highlighted in Fig. \ref{fig:intro_snaps}a,b) to cationic residues of the RBD having a greater number of close contacts with the polymers. Binding for LPGS (heparin) to the RBD residues R346 (R346), K444 (R356) and R466 (R357) are visualized in red, green, and orange, respectively. The survival probability, averaging over all binding--unbinding time series data, is shown for \textbf{(c)} LPGS and \textbf{(d)} heparin. The dashed line in panel \textbf{c} or \textbf{d} represents the double exponential fit to the data, with the value of the largest time constant $\tau$ provided in the legend.}
    \label{fig:SP}
\end{figure*}

The validity of the binding free energy obtained from the static pulling simulations depends on whether the binding--unbinding equilibrium for interactions between the charged groups of the protein and polymer has been reached within the simulation time. The time-series data shows that anionic sulfate groups of LPGS bind intermittently to various cationic residues of the RBD (see Figure~\ref{fig:SP}a).
To quantify the time required to attain binding--unbinding equilibrium, we calculate the intermittent survival probability (SP) defined as \cite{rapaport1983hydrogen} 
\begin{align}
\label{eq:SP}
    \mathrm{SP}(\tau) = \left\langle \frac{\sum_{ij}^{}s_{ij}(t)s_{ij}(t+\tau)}{\sum_{ij}^{}s_{ij}(t)}\right\rangle_{t},
\end{align}
where
\begin{align*}
\label{formula:sij}
s_{ij}(t)=\begin{cases}0&\text{if $N_{C}^{ij}(t) = 0$}\\1&\text{if  $N_{C}^{ij}(t) > 0$,}\end{cases}
\end{align*}
with $N_{C}^{ij}(t)$ being the number of close contacts (defined by an interatomic distance cutoff of \SI{3.5}{\angstrom}) between a polymer charge group $i$ and a protein residue $j$ at time $t$.
$\mathrm{SP}(\tau)$ represents the probability of finding a polymer group that is bound to a protein residue at time $t$ to be bound to the same residue at time $t+\tau$ and is shown for LPGS binding in Figure~\ref{fig:SP}c. 
The relaxation time $\tau$ refers to the largest timescale involved and is obtained by fitting a bi-exponential function ($= ae^{-t/\tau_{1}} + be^{-t/\tau_{2}} + c$) to the SP data. 
For LPGS binding, we find a relaxation time of \SI{260}{\nano\second} that is an order of magnitude smaller than the total simulation time of \SI{5}{\micro\second}, ensuring sufficient sampling. 
Moreover, we check the convergence of the free-energy profile $F(\xi)$ by splitting the simulation into five blocks, each of duration \SI{1}{\micro\second}, and calculating the free energy profile for each block. We find that the free energy landscapes converge for at least three such blocks, for LPGS binding to both the wild-type and Delta-variant RBD 
(see Figure S7 in the SI). \par{}
For heparin, we observe much slower relaxation dynamics near the RBD surface. The time-series data shows that charged groups of heparin stay bound to a single protein residue for almost the whole simulation time (Figure~\ref{fig:SP}b). From the SP function, we find a relaxation time of \SI{7.1}{\micro\second} (Figure~\ref{fig:SP}d). Thus, order of magnitude-longer simulations ($\sim\SI{50}{\micro\second}$) would be required to achieve the binding--unbinding equilibrium, which is far beyond the reach of our computational resources. \par{}
\subsection{Protein Conformational Transitions}
\begin{figure*}[h!]
    \centering
    \includegraphics[width=1\textwidth]{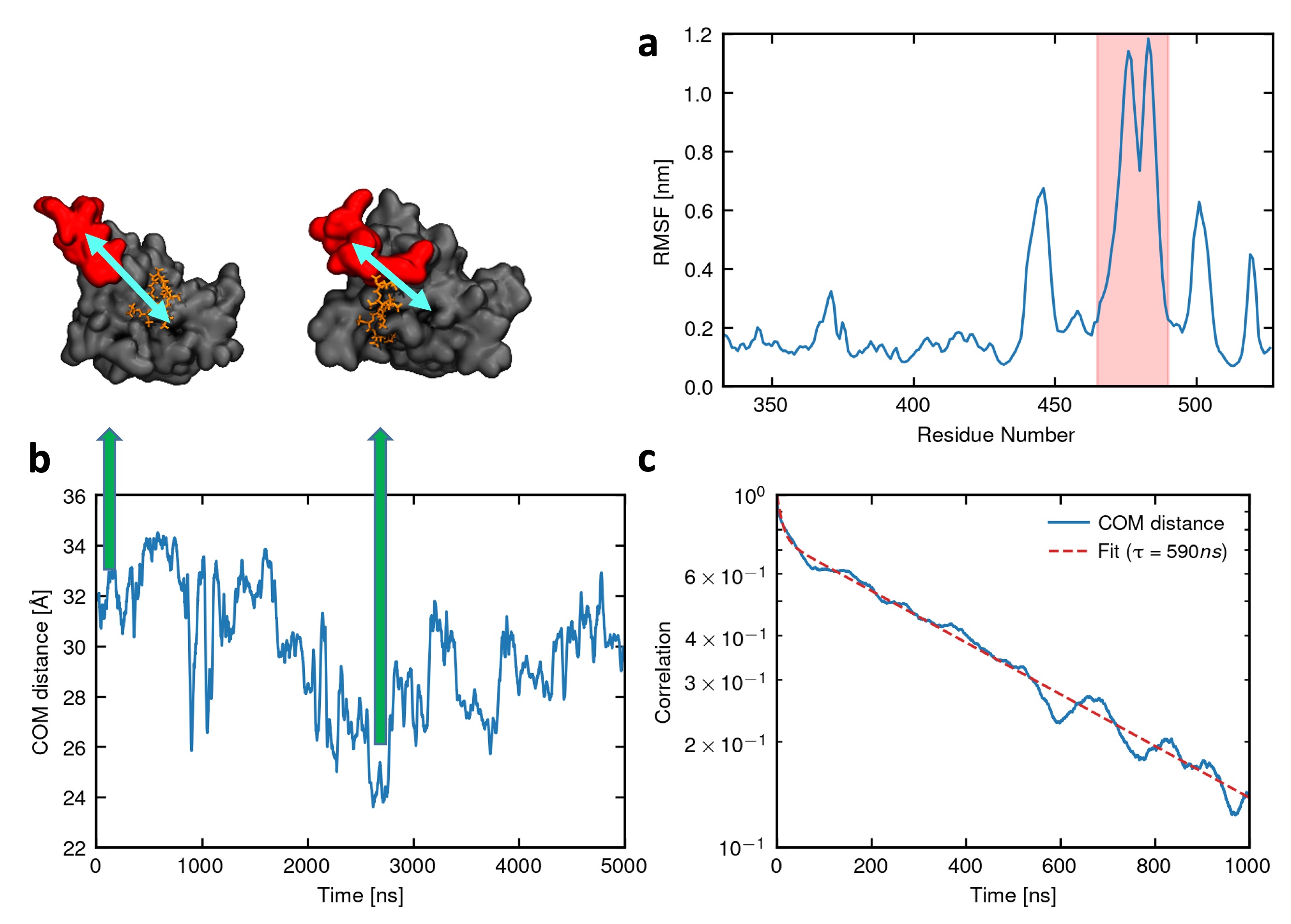}
   \caption{Conformational fluctuations of the wild-type RBD and dynamics of its loop region (residues 470 to 490) taken from static simulation with $\xi=\SI{1.6}{\nm}$\space
    \textbf{(a)} Root-mean-square fluctuation (RMSF) of the protein backbone atoms for different residues. The shaded region represents residues corresponding to the RBD's loop part. \textbf{(b)} The time series of the distance between the center-of-mass of the loop region and the remaining part of the RBD. Snapshots for two different states at the start and after 2700 ns of the simulation are displayed above. The RBD surface is shown in grey except for its loop region in red, whereas LPGS is shown in the ball-stick representation in orange. \textbf{(c)} The center-of-mass distance autocorrelation function.  The dashed line represents the double exponential fit to the data, with the value of the largest time constant $\tau$ given in the legend.
   }
    \label{fig:loop_region}
\end{figure*}
When LPGS is present near the wild-type RBD surface, we observe transitions between different protein conformations, which adds further complexity to the quantitative prediction of the binding thermodynamics. The RBD has a loop region (residues 470 to 490) that shows higher root-mean-square positional fluctuations (see Methods for the definition) and is therefore highly flexible (see Figure~\ref{fig:loop_region}a). Due to this, the loop region can switch between multiple states and thus modify the overall protein structure significantly, as seen from the time-series plot of the distance between the center-of-mass of the loop region and the remaining part of the protein in Figure~\ref{fig:loop_region}b.
From the autocorrelation function of this center-of-mass distance, we find for the RBD conformational dynamics a relaxation time of \SI{590}{\nano\second} as shown in Figure~\ref{fig:loop_region}c (for further details, see Methods). As our simulation time is an order of magnitude larger than the relaxation time, sufficient sampling of different protein configurations is ensured. However, the error in the desorption force for the Delta-variant RBD is large (see Figure~\ref{fig:simulation_results}d, e.g. at $\xi =$ \SI{0.4}{\nano\meter}) since the bound polymer attracts the flexible loop of the protein towards it 
(see snapshots in Figure S8 in the SI). \par{} 
Due to the protein structural transitions, errors for the hydration analysis can also be large (Figure \ref{fig:internal_energy}c, e.g. at $\xi =$ \SI{2.8}{\nano\meter}). Because the loop region occasionally adsorbs on the protein surface itself, multiple water molecules are released during this process 
(see Figure S9 in the SI). This accounts for large fluctuations in the number of bound water within a single simulation window, making a quantitative understanding of the entropic contribution due to water release difficult.  
Note that as the loop region is part of the receptor binding motif that forms direct contact with the host cell receptor protein ACE2, its flexibility might help in adapting the viral spike protein structure for binding to other cell receptors and thus improve viral infectivity. \par{}
%
\section{Conclusions} 
We demonstrate a method to obtain the binding free energy of long polymers (\num{10}--\SI{100}{\kilo\dalton}) typically considered in experiments, from the simulated free-energy profile of shorter polymer desorption from a protein. This requires correctly accounting for (i) the binding volume of the polymer, (ii) the polymer stretching free energy, and (ii) the \textit{avidity entropy} due to the different possible ways the polymer can bind to the protein (cf. Eq.~\eqref{eq:f_bind_long}). We validate our method by favorable comparison with the experimental free-energy of binding between LPGS and the wild-type SARS-CoV-2 spike protein RBD and reproducing accurately the polymer-length dependence. We find that anionic LPGS binds more strongly
to the delta-variant RBD (with extra 2 mutated cationic residues) than the wild-type RBD, underlining the role of electrostatic interactions. The LPGS--RBD binding at $T = 300$ K is found to be enthalpy driven, although a large enthalpy--entropy compensation is observed. Decomposing the enthalpy of binding, we find a huge cancellation between the direct polymer--protein interaction energy contribution and the indirect solute--solvent interaction energy and solvent reorganization energy contributions. These observations signify the importance of solvent and entropic effects in molecular binding \cite{sahoo2022role, Method_article, malicka2022interaction}. \par{}
We identify a highly flexible loop region of the RBD, which transitions between different states with a relaxation time of ~\SI{600}{\nano\second}. Thus, slow protein conformation transitions 
can add complications in predicting the binding thermodynamics accurately. Moreover, we show that modeling the adsorption of a highly charged polymer, e.g., the drug heparin to RBD, requires a high computational effort as the relaxation time for the binding equilibrium between their charged groups is $\sim \SI{7}{\micro\second}$. \par{}
Comparing three different simulation protocols for obtaining the polymer adsorption free energy difference, we find that the extrapolation method using dynamic pulling data gives larger values than umbrella sampling or static pulling results. 
This hints at the relevance of chain reconfiguration effects when pulling a polymer from an absorbing surface, as is relevant in force spectroscopy experiments \cite{ geisler2010controlling} and biological nonequilibrium scenarios.  
Thus, the dynamic pulling method is not only useful in generating initial configurations for the static pulling method but also for understanding friction and diffusion in the bound protein--polymer complex, which is important for kinetics of the binding process. \par{}
\section{Methods}
\subsection{MD Simulations}
\textbf{\textit{Models, Parameters, and Simulation Set-Up.}}
The coordinates for the wild-type RBD of the SARS-CoV-2 spike protein are obtained from the deposited crystal structure (PDB ID: 6M0J) \cite{lan2020structure}. The Delta-variant RBD with L452R and T478K mutations is built using PyMOL. The structure of the heparin pentamer is built using CHARMM-GUI Glycan Reader \& Modeler \cite{jo2008charmm, brooks2009charmm}. The structure of LPGS undecamer is built using Avogadro software \cite{hanwell2012avogadro}. CHARMM36m \cite{huang2017charmm36m} and CHARMM Carbohydrates \cite{Charmm_carbo} force field parameters are used to model the protein and heparin, respectively. 
Parameters and partial atomic charges for LPGS are modeled with the CHARMM General force field \cite{vanommeslaeghe2010charmm,yu2012extension} and obtained using the CGenFF program \cite{vanommeslaeghe2012automation_1,vanommeslaeghe2012automation_2}. 
CHARMM-compatible TIP3P water \cite{jorgensen1983comparison,mackerell1998all} and ion parameters \cite{venable2013simulations} are used. RBD/LPGS and RBD/heparin are arranged and solvated in boxes of sizes $7\times7\times \SI{9.5}{\nano\meter^3}$ and $7\times7\times\SI{10}{\nano\meter^3}$, respectively. Enough $\mathrm{Na^{+}}$ ions are added to charge neutralize each system, then $\mathrm{Na^{+}}$ / $\mathrm{Cl^{-}}$ ion pairs are added to obtain a \SI{150}{\milli\Molar} NaCl solution estimated from the mole fraction of ion pairs and water. \par{}
Before starting pulling simulations, unconstrained simulations are performed for at least \SI{1}{\micro\second} in the $NpT$ ensemble at $T =$ \SI{300}{\kelvin} and $p =$ \SI{1}{\bar} with periodic boundary conditions in xyz directions, using the GROMACS 2020.6 package \cite{abraham2015gromacs}. During the simulations, backbone atoms of three residues of the RBD are fixed to stop its center-of-mass translation and its rotation around the principal axes. The stochastic velocity rescaling thermostat \cite{bussi2007canonical} with a time constant of $\tau = \SI{0.1}{\pico\second}$ is used to control the temperature, while for the pressure control an isotropic Parrinello-Rahman barostat \cite{parrinello1981polymorphic} is used with a time constant of $\tau = \SI{2}{\pico\second}$ and a compressibility of $\kappa= \SI{4.5e-5}{\per\bar}$. 
The LINCS algorithm \cite{hess2008p} is used to constrain the bonds involving H-atoms, allowing a timestep of $\Delta t = \SI{2}{\femto\second}$. Electrostatic interactions are computed using the particle mesh Ewald 
method \cite{darden1993particle} with a real-space cutoff distance of \SI{1.2}{\nano\meter}, while van der Waals interactions are modeled using Lennard-Jones potentials with a cutoff distance of \SI{1.2}{\nano\meter} where the resulting forces smoothly switch to zero between of \SI{1}{\nano\meter} to \SI{1.2}{\nano\meter}. A data saving frequency of \SI{100}{\pico\second} is used. \par{} 
\textbf{\textit{Dynamic and Static Pulling Simulations.}}
For performing pulling simulations, the z-distance between the center-of-mass of the protein and one terminal atom of the polymer is chosen as a reaction coordinate $\xi$, which is shifted such that $\xi=0$ when the terminal atom bound to the surface. 
Constant velocity pulling simulations (which we refer to as dynamic pulling)  are conducted in the $NVT$ ensemble with an umbrella potential and a spring constant $k =$ \SI{1660}{\pico\newton\per\nano\meter}. In a dynamic pulling simulation, a spring attached to a terminal polymer atom is pulled with a constant velocity $v$, and the pulling force $f$ is obtained from the extension of the spring from its equilibrium position. The used pulling speed $v$ ranges from \SI{0.006}{\meter\per\second} to \SI{1.2}{\meter\per\second}. To ensure that the whole polymer is desorbed from the protein surface, a simulation time in the range of \SI{5}{\nano\second} to \SI{1000}{\nano\second} is used so that at the end, a pull distance of roughly \SI{6}{\nano\meter} is reached. For each pulling speed, simulations are repeated until the combined simulation time reaches \SI{1000}{\nano\second} or more. The coordinates of the system are saved every \SI{100}{\pico\second} and pulling forces are recorded every \SI{100}{\femto\second}. \par{}
For the static simulations, the starting configurations are generated from multiple configurations from the dynamic pulling simulations with a \SI{0.4}{\nano\meter} spacing of the reaction coordinate. The time for each static simulation window is set to \SI{5}{\micro\second}, summing up to a total simulation time of \SI{45}{\micro\second}.
The free-energy profile from the static simulations is calculated by first computing the average force for each simulation window using the last \SI{4.5}{\micro\second} data and then integrating over the average force profile, see Eq.~\eqref{formula:Path_integral_f}.
\par{}
\textbf{\textit{Umbrella Sampling Simulations.}}
Taking configurations from the dynamic pulling simulations for the wild-type RBD--LPGS system, we have conducted 30 simulations for \SI{100}{\nano\second} each where umbrella or harmonic potentials are applied
to restrain the system at different values of the reaction coordinate $\xi$ (the same as defined before), from  \SIrange{0.1}{3.0}{\nano\meter} with a window spacing of \SI{0.1}{\nano\meter}. 
The same spring constant as in the dynamic pulling is used for each umbrella window. The weighted histogram analysis method \cite{kumar1992weighted}, implemented in the GROMACS module \textit{wham} \cite{abraham2015gromacs}, is used to obtain the free-energy landscape $F({\xi})$ for LPGS desorption from the RBD surface, shown in Figure \ref{fig:simulation_results}f. The first \SI{20}{\nano\second} data for each umbrella window is discarded for the calculation of $F({\xi})$. 
The overlapping of histograms for consecutive umbrella windows needed for the accurate computation of $F({\xi})$ and its convergence check by taking different lengths of the simulation data are shown in Figure S10 in the SI. \par{}
\textbf{\textit{Constant-Force Stretching of LPGS.}}
An LPGS undecamer, placed in a rectangular simulation box of size $5 \times 5 \times \SI{9}{\nano\meter^3}$, charge-neutralized by adding counterions ($\mathrm{Na^{+}}$) and solvated with water in a \SI{150}{\milli\Molar} NaCl solution, is taken for the simulations. The simulation-related parameters coincide with the ones of the unconstrained RBD/LPGS simulation in the $NpT$ ensemble except, that the GROMACS 2021.3 package \cite{abraham2015gromacs} and ion parameters of Loche et al.\ \cite{locheTransferableIonForce2021} are used. Additionally, we have chosen a higher trajectory saving frequency of \SI{10}{\pico\second}. The equilibration process starts with employing the steepest descent algorithm for the initial energy minimization. This is followed by two stages of simulation: a \SI{500}{\pico\second} $NVT$ simulation and a \SI{2}{\nano\second} $NpT$ simulation, during which the polymer's particles' positions are restrained. \par
\begin{figure}[h]
    \centering
    \includegraphics[width=0.49\textwidth]{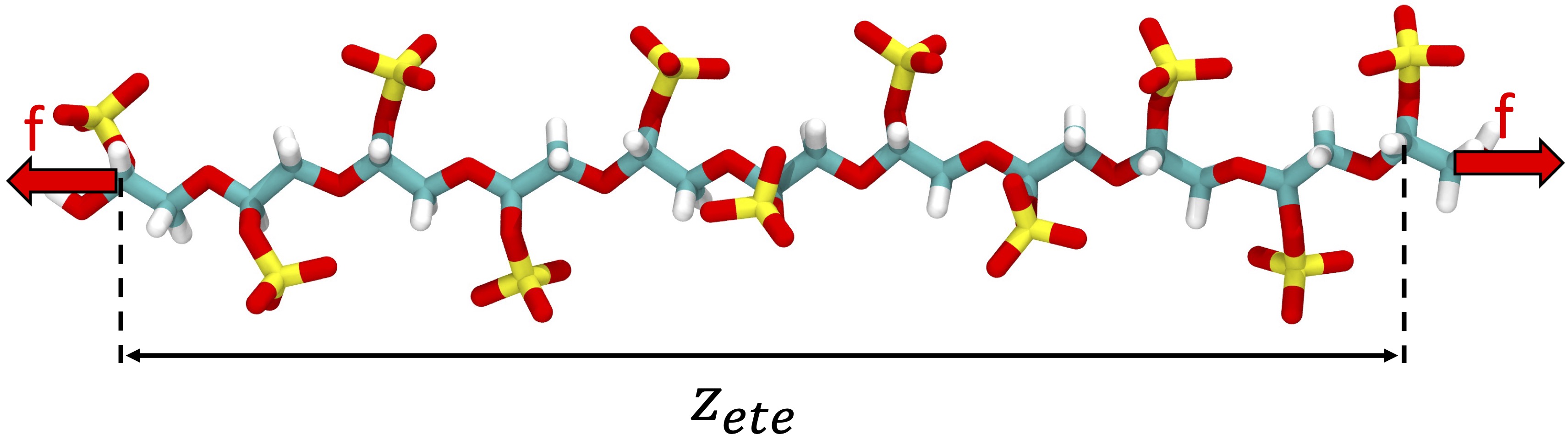}
    \caption{Stretching protocol showing the LPGS undecamer at applied forces $f =$ \SI{800}{\pico\newton}. Color coding for the different atom types: hydrogen in white, carbon in cyan, oxygen in red, and sulfur in yellow.}
    \label{fig:lpgs_stretching}
\end{figure}

{}
To determine the stretching free-energy $\Delta F_{\mathrm{stretch}}$ of LPGS, we have performed in the $NpT$ ensemble several production simulations in each of which a constant force (for forces between \SI{1}{\pico\newton} and \SI{1000}{\pico\newton}) is applied to the polymer ends in opposite directions along the z-axis. 
The anchor points for the constant force are namely the first (C1) and last (C22) carbon atoms of the LPGS undecamer along its backbone as depicted in Figure~\ref{fig:lpgs_stretching}. The average extension $\langle z_{ete} \rangle$ for ten monomers, defined as the distance between the carbon atoms C1 and C21 in the pulling direction, is measured. 
This ensures the inclusion of all relevant monomers under strain.
We have performed $NpT$ production simulations for different durations from 200 ns (at higher forces) to 2000 ns (at lower forces), depending on the relaxation time for the ion distribution around the polymer and for the polymer's end-to-end distance at different applied forces. For data analysis, the initial part (\SI{20}{\nano\second} to \SI{50}{\nano\second} depending on the applied force) of a production run is discarded for the equilibration which accounts for initial stretching or shortening of LPGS. \par{}
%
\subsection{Simulation Data Analysis}\label{subsec:sim_analysis}
Simulation data are visualized and analyzed with VMD \cite{humphrey1996vmd} and the software package MDAnalysis \cite{gowers2016mdanalysis,michaud2011mdanalysis}, respectively.\par{}
\textbf{\textit{The} iPFRC \textit{Model and Stretching Free Energy of LPGS.}} \label{subsubsec:U_stretch}
The force $f$ versus extension $z_\mathrm{ete}$ profile obtained from constant-force stretching of LPGS is depicted in 
Figure~\ref{fig:stretching_free_Energy}a (for the complete range of forces used in this study, the profile is shown in 
Figure S11 in the SI). 
To interpolate the data points, we use the heuristic force--extension relation of the iPFRC model \cite{hankeStretchingSinglePolypeptides2010}:
\begin{equation}\label{eq:f-e_iPFRC}
    f = \frac{\si{\kT} z_{\mathrm{ete}}}{L} \left( 
            \frac{3}{a_{\mathrm{Kuhn}}}
            + \frac{1}{c a_0} \frac{z_{\mathrm{ete}} / L}{1 - z_{\mathrm{ete}} / L}
    \right).
\end{equation}
Here, $L$ is the contour length of the polymer, and $a_0$ is the equilibrium monomer length. The Kuhn length $a_{\mathrm{Kuhn}}$ is defined by the linear stretching response at low applied forces and $c$ is a free parameter, whose choice accounts for restricted backbone dihedral rotation and side chain interactions. The iPFRC model has been shown to describe the force--extension relation of various polypeptides quite well \cite{schwarzlForceResponsePolypeptide2020}, when a force-dependent contour length $L(f)$ is introduced additionally, 
\begin{equation}
    L(f) = L_0 \left(1 + \frac{\sqrt{\gamma_1^2 + 4\gamma_2 f} -\gamma_1}{2 \gamma_2} \right). 
\end{equation}
$L_0$ denotes the unstretched contour length of the polymer. The linear stretching modulus $\gamma_1$ and the nonlinear coefficient $\gamma_2$ describe the force-dependent extension of a monomer at zero temperature in vacuum. The variables $a_0$, $a_{\mathrm{Kuhn}}$, $c$, $\gamma_1$, and $\gamma_2$ have been used as free fitting parameters and their values corresponding to the shown line in Figure~\ref{fig:stretching_free_Energy}a are reported in Table~\ref{tab:iPFRC}. 
The stretching free energy $\Delta F_{\mathrm{stretch}}$ follows by integrating the fitted force--extension curve 
Eq. \ref{eq:u_stretch} and is shown in Figure~\ref{fig:stretching_free_Energy}b. The choice of setting the lower bound of the integral in Eq. \ref{eq:u_stretch} to zero is grounded on the premise that, in the absence of an applied force, 
the expected value of the average extension is zero.
\begin{table}[h]
    \centering
     \caption{Fitting parameters of the iPFRC model}
    \begin{tabular}{|c|c|}
           \hline
       Parameter & Value  \\
       \hline
        $a_0$ & \SI{366.82+-0.67}{\pico\meter}\\ 
        \hline
        $a_{\mathrm{Kuhn}}$ & \SI{0.873+-0.033}{\nano\meter}\\
        \hline
        $c$ & \SI{0.793+-0.018}{}\\
        \hline
        $\gamma_1$ & \SI{96+-36}{\nano\newton}\\
        \hline
        $\gamma_2$ & \SI{500+-2000}{\nano\newton}\\
      \hline
    \end{tabular}
    \label{tab:iPFRC}
\end{table}
\textbf{\textit{Internal Energy Decomposition.}}
Interaction energy calculations are done using the GROMACS module \textit{energy} \cite{abraham2015gromacs}. The average energy is calculated for the adsorbed state,  $\xi =$ \SI{0}{\nano\meter}, and the desorbed state, $\xi =$ \SI{3.6}{\nano\meter}, for interactions between different components of the system: protein, polymer, and solvent (water and ions), and the energy differences are computed. Only the short-range part of the Coulomb interaction with a cutoff distance 1.2 nm is included in these calculations, as the full, long-range electrostatic energy of a subsystem with a non-zero net charge with periodic boundary conditions diverges. However, calculation of the net change in interaction energy, $\Delta U_\mathrm{b}$, 
for the whole system includes also the long-range electrostatic contribution. Simulation data of \SI{2.5}{\micro\second} (25000 frames) and \SI{5}{\micro\second} (50000 frames) are used for calculating the interaction energies for adsorbed and desorbed states, respectively. \\ \par{}
\textbf{\textit{Distance Criteria for Close Contacts and Bound Water and Ions.}}
For calculating the number of close contacts, we define a contact by an atom of LPGS falling within \SI{3.5}{\angstrom} of any atom of a protein residue. 
The same cutoff distance (of \SI{3.5}{\angstrom}) criterion is used to calculate the protein-bound water molecules and ions. \par{}
\textbf{\textit{Root-Mean-Square Fluctuation of the RBD Structure.}}
The root-mean-square fluctuation (RMSF) in positions of the backbone atoms for each RBD residue is calculated using the GROMACS module \textit{rmsf} \cite{abraham2015gromacs} and the following formula.
\begin{align}
    \mathrm{RMSF}(i) = \sqrt{\frac{1}{N_t}\displaystyle \sum_{t=1}^{N_t} (\textbf{r}_{i}(t)-\textbf{r}_{i}^\mathrm{ref})^2},
\end{align}
where $N_t$ is the total number of timesteps, $\mathbf{r}_{i}(t)$ is the position of residue $i$ at time $t$, and $\mathbf{r}_{i}^\mathrm{ref}$ is the position the same residue in the reference state (the native state of RBD). \par{}
\textbf{\textit{Relaxation Time for the RBD's Loop Region Movement.}}
The movement of the flexible loop region (residues 470 to 490) of the RBD is tracked by calculating the center-of-mass (COM) distance between the loop region and the rest of the protein. To get an estimate of the relaxation time, we calculate the COM distance autocorrelation function $C(t)$ defined as: 
\begin{align}
\label{formula:acf}
    C(t) =  \frac{\left\langle (A(t')-\overline{A})(A(t'+t)-\overline{A})\right\rangle_{t'}}{\left\langle (A(t')-\overline{A})^2\right\rangle_{t'}},
\end{align}
where $A(t')$ is an observable at an initial time $t'$, $\overline{A}$ is the time average of observable $A$, $\langle \cdot \rangle_{t'}$ represents the time-origin averaging. The relaxation time is estimated by fitting a bi-exponential function ($= ae^{-t/\tau_{1}} + be^{-t/\tau_{2}}$) to the COM distance autocorrelation $C(t)$ and refers to the longest timescale involved ($\tau_2$). \par{}
\textbf{\textit{Error Estimations.}}
Errors for the static pulling, the average extension of LPGS, internal energy calculations, and the number of protein-bound water molecules and ions are estimated by using the block averaging method by Flyvbjerg and Petersen \cite{flyvbjerg1989error}. The number of blocks is changed until the standard error of the different blocks converges to a constant value. In case the standard error does not converge, the maximum standard error is used as an error estimate. For the dynamic pulling, the error is estimated by calculating the standard error of $\Delta F$ for each pulling rate. \par{}
%
\section{Acknowledgements}
We would like to acknowledge funding by IRTG GRK2662 and ERC NoMaMemo and computing time on
the HPC cluster at the Physics department of Freie Universit\"at Berlin. \par{}
\section{Author Contributions}
A.K.S. and R.R.N. designed research; L.N., C.H., and A.K.S. performed research; L.N., C.H., and A.K.S. analyzed data; and A.K.S. and R.R.N. wrote the paper with inputs from all authors. \par{}
\begin{suppinfo}
A derivation of the standard free-energy of binding from a polymer desorption free-energy profile; Additional figures.
\end{suppinfo}
%

\providecommand{\latin}[1]{#1}
\makeatletter
\providecommand{\doi}
  {\begingroup\let\do\@makeother\dospecials
  \catcode`\{=1 \catcode`\}=2 \doi@aux}
\providecommand{\doi@aux}[1]{\endgroup\texttt{#1}}
\makeatother
\providecommand*\mcitethebibliography{\thebibliography}
\csname @ifundefined\endcsname{endmcitethebibliography}
  {\let\endmcitethebibliography\endthebibliography}{}

\end{document}


\clearpage

\section{SI Text}
\subsection{Standard Free-Energy of Binding from a Polymer Desorption Free-Energy Profile}
\label{subsec:binding_fg}
\begin{figure}[h]
\centering
\includegraphics[width=0.51\textwidth]{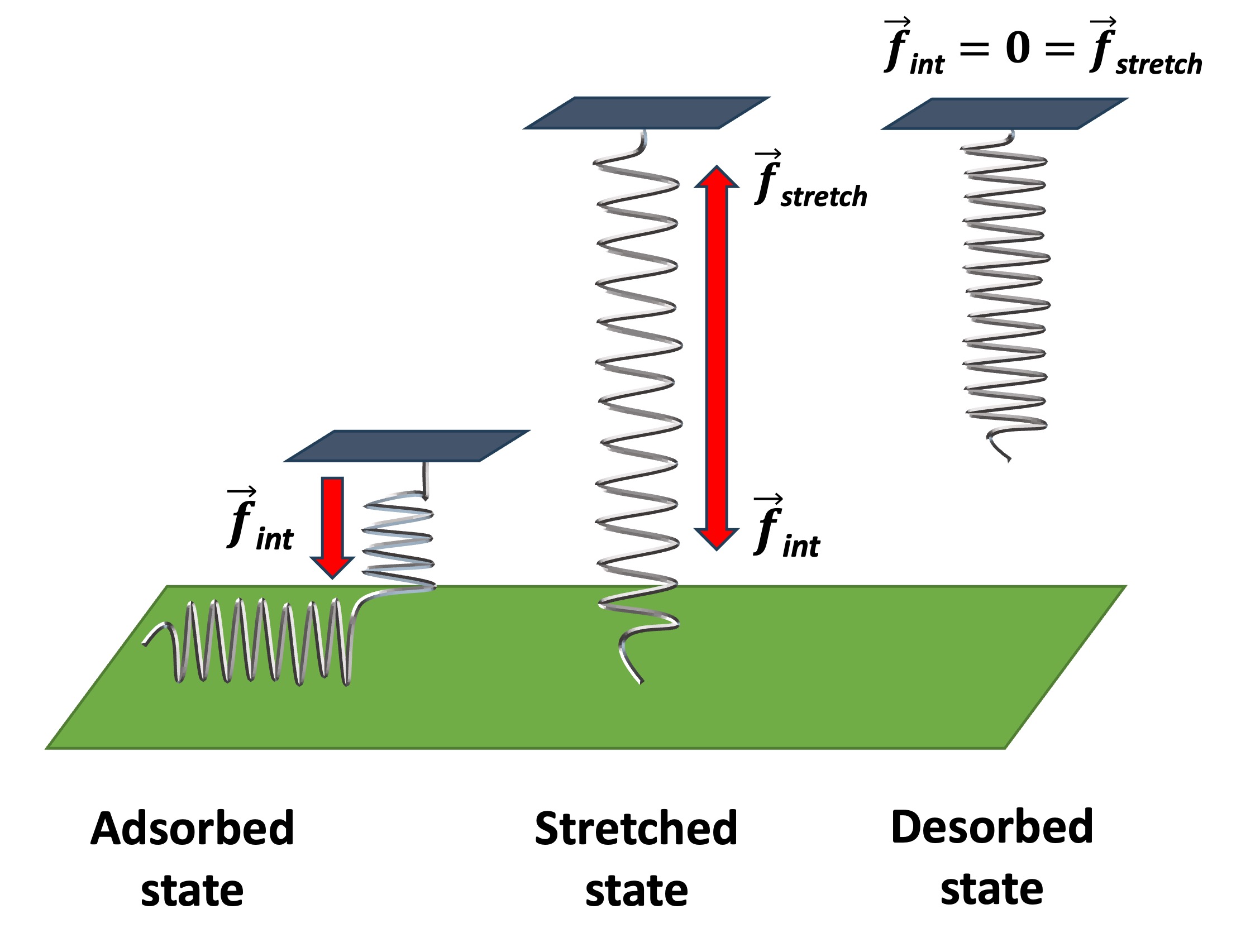}
\caption{The pulling simulation protocol leads to the additional stretching free energy term $\Delta F_{\mathrm{stretch}}$ in Eq. \ref{eq1}.}
\label{fig:stretch_concept}
\end{figure}

\begin{figure}[h]
\centering
\includegraphics[width=0.58\textwidth]{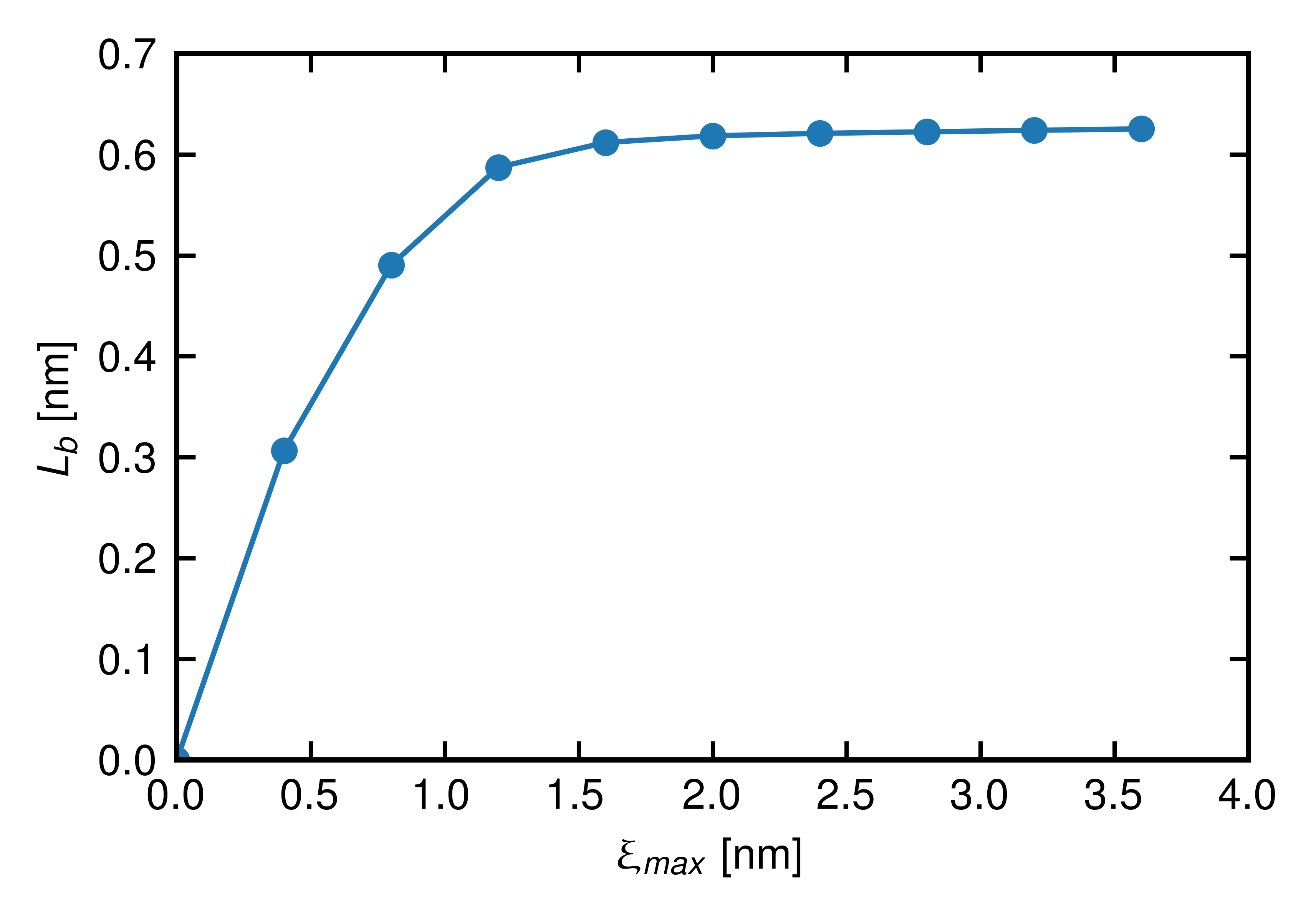}
\caption{The ligand binding length $L_\mathrm{b}$ in Eq. \ref{eq4} as a function of the maximum pulled distance $\xi_\mathrm{max}$ taken to define the bound region.
}
\label{fig:Lb_vs_xi}
\end{figure}

\begin{figure}[h!]
\begin{subfigure}{.58\textwidth}
  \centering
  \includegraphics[width=1.0\linewidth]{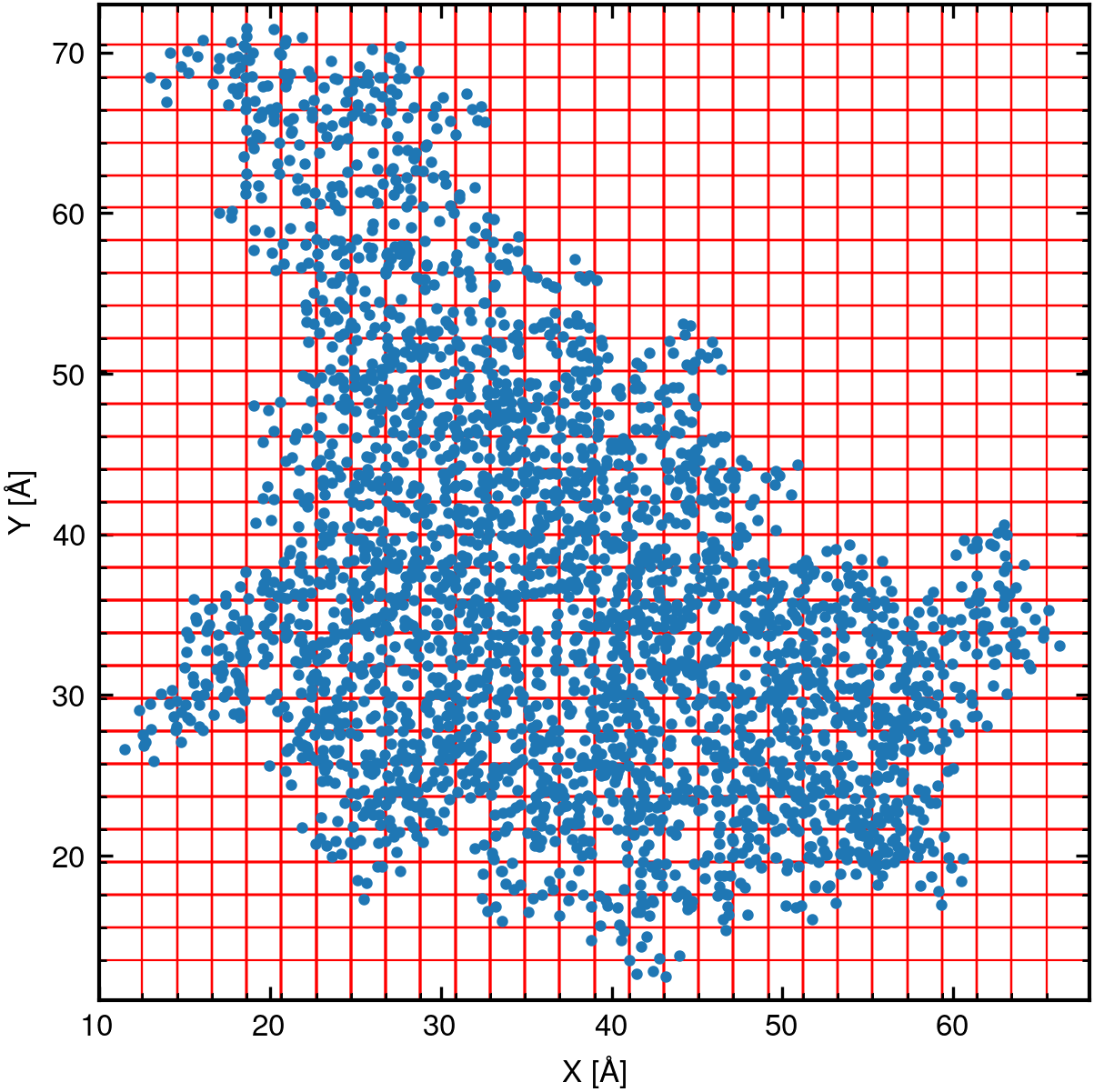}
\end{subfigure}%
\caption{Calculation of the area $A_\mathrm{pro}$ of the protein  projected on the xy plane. The simulation box in the xy plane is divided into 850 square pixels with pixel width 2 \AA{} and the area elements of the pixels containing at least one protein atom (blue dots) are summed to yield the total area $A_\mathrm{pro} = 18.1\ \mathrm{nm}^2$.}
\label{fig:A_pro}
\end{figure}

The standard binding free-energy $\Delta F_\mathrm{b}^0$ of a polymer to a protein as a function of the degree of polymerization $N$ can be obtained from simulations of a shorter polymeric unit $N_\mathrm{sim}$  as 
\begin{equation}\label{eq1}
    \Delta F_\mathrm{b}^0(N) = \Delta F_\mathrm{b}(N_\mathrm{sim}) + \Delta F_\mathrm{V} -\Delta F_\mathrm{stretch}(N_\mathrm{sim}) -T\Delta S_\mathrm{avidity}(N, N_\mathrm{sim}) .
\end{equation}
The avidity entropy contribution $\Delta S_\mathrm{avidity} \approx k_\mathrm{B} \ln(N/N_\mathrm{sim})$ is only present for $N>N_\mathrm{sim}>n_\mathrm{b}$, with $n_\mathrm{b}$ being the number of binding sites on the protein, as explained in the main text.
The polymer stretching free-energy $\Delta F_{\mathrm{stretch}}$ is subtracted, as the stretching effect arising from our simulation protocol (see Figure~\ref{fig:stretch_concept}) is absent in experimental measurements of equilibrium binding, and furthermore, $\Delta F_\mathrm{b}^0$ should be independent of the reaction coordinate used in simulations. 
$\Delta F_\mathrm{V}(N_\mathrm{sim}) = - k_\mathrm{B}T \ln(V_\mathrm{u}/V_\mathrm{0})$ represents the free-energy change for transforming from the standard-state volume $V_0$ ($= 1.661 \mathrm{nm}^3$), corresponding to a 1 M concentration, to the sampled volume $V_\mathrm{u}$ of the unbound state in simulations.
The free-energy of binding $\Delta F_{\mathrm{b}}$ can be obtained from the polymer--protein interaction free-energy profile $F(\textbf{r})$ as:
\begin{equation}\label{eq2}
    \Delta F_\mathrm{b} = F_\mathrm{b} - F_\mathrm{u} = -k_\mathrm{B}T \ln{\left(\frac{Z_\mathrm{b}}{ Z_\mathrm{u}}\right)} 
    = 
    -k_\mathrm{B}T \ln{\left(\frac{\int_{\mathrm{b}} d^3\textbf{r} \; e^{-F(\textbf{r})/k_\mathrm{B}T}}{\int_{\mathrm{u}} d^3\textbf{r} \; e^{-F(\textbf{r})/k_\mathrm{B}T}} \right)},
\end{equation}
where $k_\mathrm{B}$ is the Boltzmann constant, $T$ is the temperature, and $Z_\mathrm{b}$ and $Z_\mathrm{u}$ are the partition functions of the bound and unbound regions, respectively.
As we have a one-dimensional desorption free-energy profile $F(\xi)$, as shown in Figure 2e in the main text, 
and it is flat in the desorbed state, subtracting the polymer desorption free energy $\Delta F = F(\xi \to \infty)$ for setting $F(\xi) = 0 $ in the unbound region implies 
\begin{align}\label{eq3}
     \Delta F_\mathrm{b} = -k_\mathrm{B}T \ln \left( \frac{\int_0^{\xi_\mathrm{max}} d\xi \; A(\xi) e^{-(F(\xi) -\Delta F)/k_\mathrm{B}T}}{V_\mathrm{u}} \right),
\end{align}
where $V_\mathrm{u} = \int_{\mathrm{u}} d^3\textbf{r}$ is the volume of the unbound state, and $F(\xi) -\Delta F < 0$. 
Approximating the area $A(\xi)$ available in the bound state along the orthogonal directions to the reaction coordinate $\xi = z$ with the area of the protein projected on the xy plane, $A(\xi) = A_\mathrm{pro}$, and defining the ligand binding length as 
\begin{equation}\label{eq4}
    L_\mathrm{b} = \int_0^{\xi_\mathrm{max}}  d\xi \; e^{-F(\xi)/k_\mathrm{B}T},
\end{equation}
we obtain from Eq. \ref{eq3} 
\begin{equation}\label{eq5}
    \Delta F_\mathrm{b} = -\Delta F - k_\mathrm{B}T \ln{\bigg(\frac{A_\mathrm{pro} L_\mathrm{b}}{V_\mathrm{u}}\bigg)} = -\Delta F - k_\mathrm{B}T \ln{\bigg(\frac{V_\mathrm{b}}{V_\mathrm{u}}\bigg)},
\end{equation}
where $V_{b}$ is the ligand binding volume. Note that the value of $L_\mathrm{b}$ is rather insensitive to the cutoff value $\xi_\mathrm{max}$ taken to define the bound region, as lower $F(\xi)$ values predominantly contribute to the integral in Eq. \ref{eq4}, see Figure \ref{fig:Lb_vs_xi}. $A_\mathrm{pro}$ is obtained numerically as described in Figure \ref{fig:A_pro}.
%
Inserting the expressions for $\Delta F^{0}_\mathrm{b}$ from Eq. \ref{eq5}, $\Delta F_\mathrm{V}$, and $\Delta S_\mathrm{avidity}$ in Eq. \ref{eq1}, we obtain  
\begin{equation}\label{eq6}
    \Delta F_\mathrm{b}^0(N) = -\Delta F(N_\mathrm{sim}) -k_\mathrm{B}T\ln{\bigg(\frac{V_\mathrm{b}(N_\mathrm{sim})}{V_\mathrm{0}}\bigg)} -\Delta F_{\mathrm{stretch}}(N_\mathrm{sim}) -k_\mathrm{B}T\ln{\bigg(\frac{N}{N_\mathrm{sim}}\bigg)} ,
\end{equation}
where the sampled volume $V_\mathrm{u}$ of the unbound state cancels out. The above final expression is reproduced in the main text as Eq. 6.
\par{}

\clearpage
\section{Additional Figures}
\begin{figure}[H]
    \centering
    \includegraphics[width=0.65\textwidth]{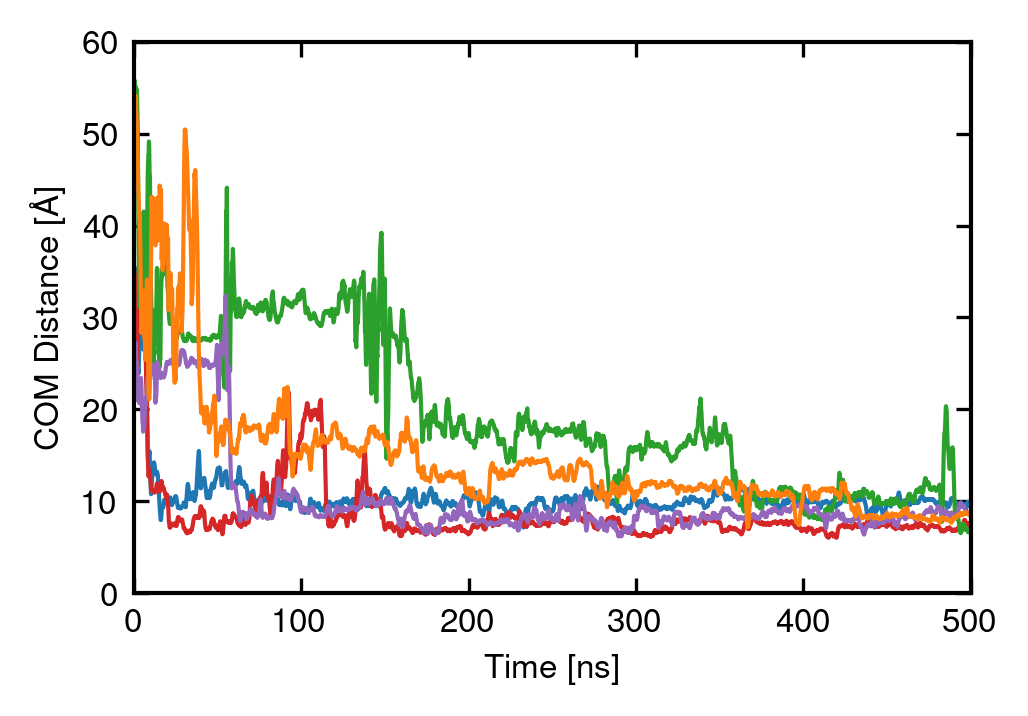}
    \caption{Time series of the distance between the center-of-mass of LPGS and the cationic patch (residues 346, 355, 356, 357 and 466) of the wild-type RBD shown for five independent simulations with different starting positions of LPGS w.r.t. the RBD.} 
    \label{fig:Adsorption_sim}
\end{figure}

\begin{figure}[H]
    \centering
    \includegraphics[width=0.6\textwidth]{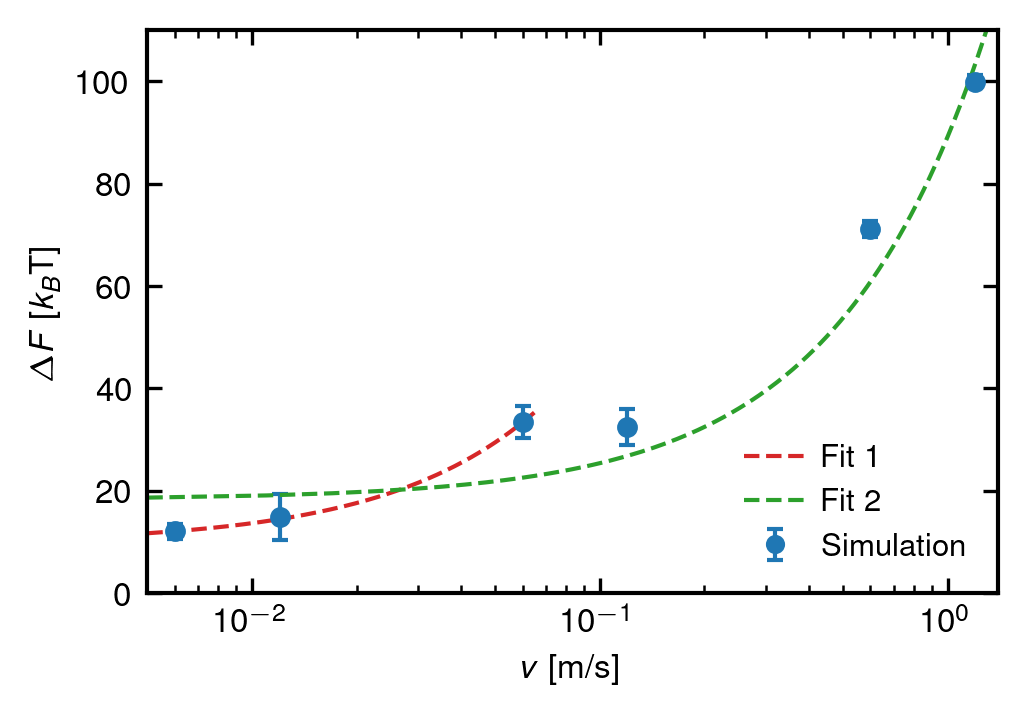}
    \caption{The equilibrium free energy $\Delta F (v=0)$ of LPGS desorption from the wild-type RBD surface obtained from a linear extrapolation in the pulling velocity $v$ of the dynamic pulling simulation values to $v=\SI{0}{\meter\per\second}$. Fit 1 represents the linear extrapolation using data for the lowest three pulling velocities (also shown in Figure 
    2c in the main text), whereas Fit 2 represents that using all data points. Note the log scale for $v$.
    $\Delta F (v=0)$ from Fit 1 and Fit 2 are $9.7 \pm 1.6\ k_\mathrm{B}T$ and $18.3 \pm 1.1\ k_\mathrm{B}T$, respectively. The error in $\Delta F (v=0)$ is from the least square fitting of the data including their error.} 
    \label{fig:extrapolation_high}
\end{figure}

\begin{figure}[H]
    \centering
    \includegraphics[width=0.95\textwidth]{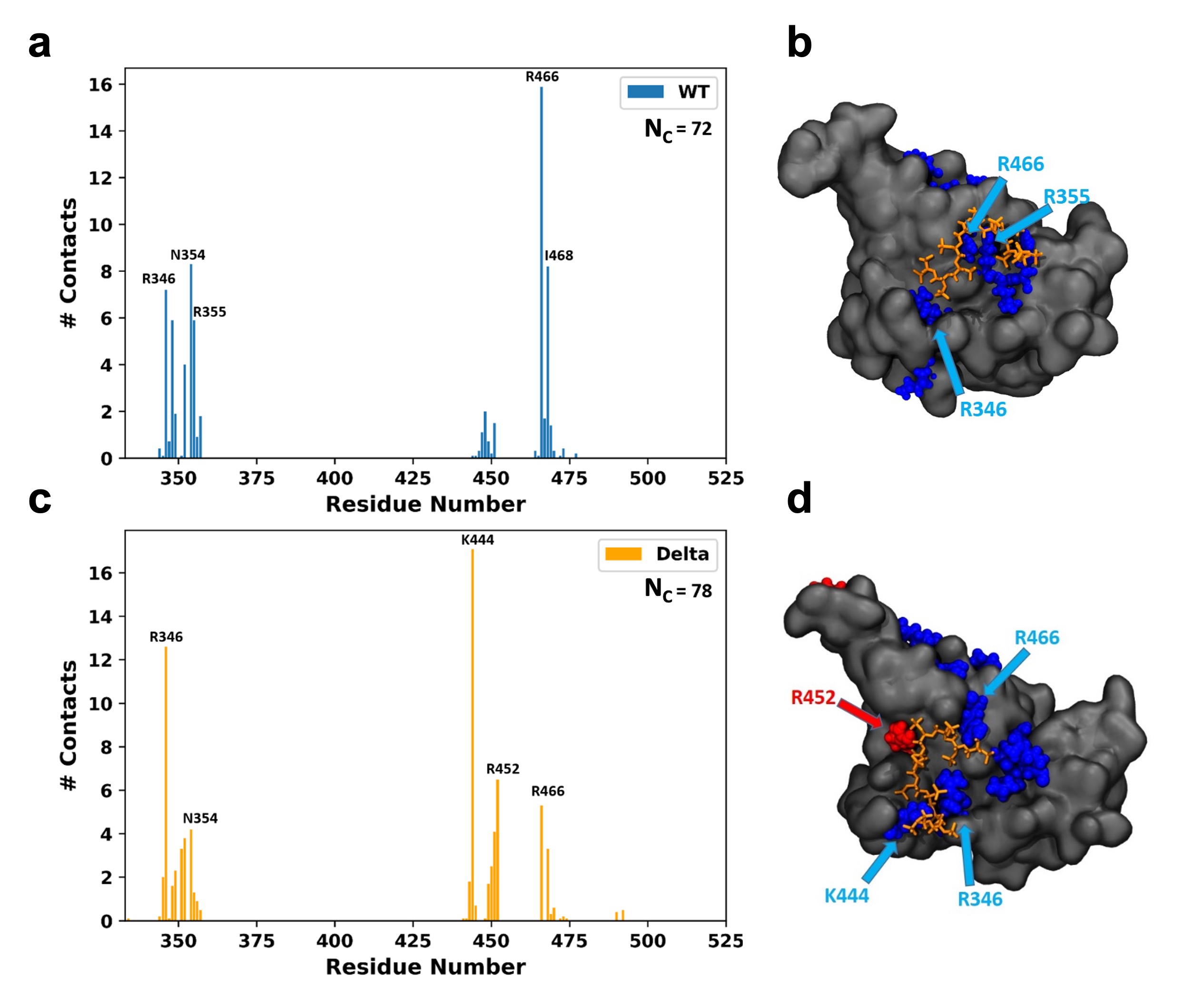}
    \caption{The average number of close contacts of LPGS with different protein residues observed in the \SI{5}{\micro\second}-long simulation at a pulled distance $\xi =$ \SI{0.8}{\nano\meter} of the LPGS terminus from the RBD surface for \textbf{(a)} the wild-type and \textbf{(c)} the Delta-variant. Representative simulation snapshots for LPGS interactions with \textbf{(b)} the wild-type RBD and \textbf{(d)} the Delta-variant RBD. LPGS is shown in the ball--stick representation in orange, whereas the protein surface is rendered in grey. Cationic residues of the protein are pointed out in blue and the additional two mutated residues in red, the residues having a substantial number of contacts with LPGS are indicated.
    }
    \label{fig:close_contacts}
\end{figure}

\begin{figure}[H]
    \centering
    \includegraphics[width=1\textwidth]{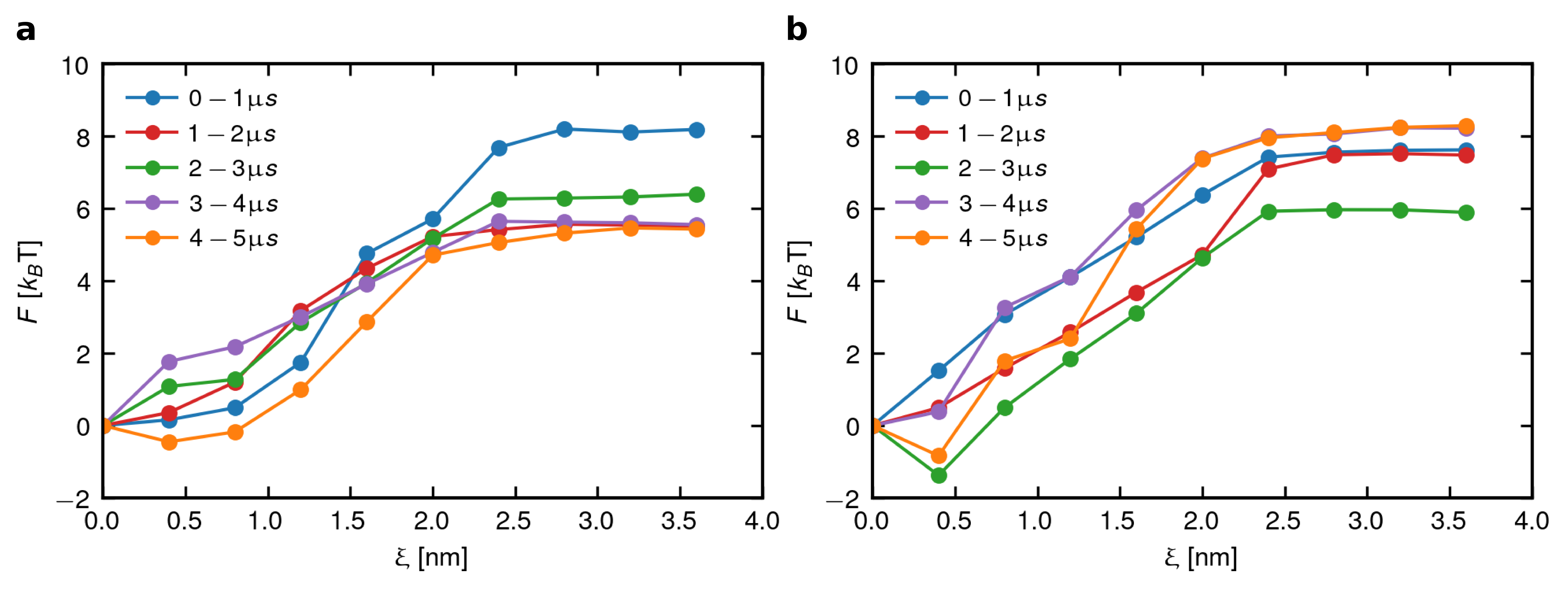}
    \caption{Free energy profiles obtained using data from different blocks (time intervals) of the static pulling simulations for the desorption of LPGS from \textbf{(a)} the wild-type RBD and \textbf{(b)} the Delta-variant RBD.} 
    \label{fig:blocks}
\end{figure}

\begin{figure}[H]
    \centering
    \includegraphics[width=0.6\textwidth]{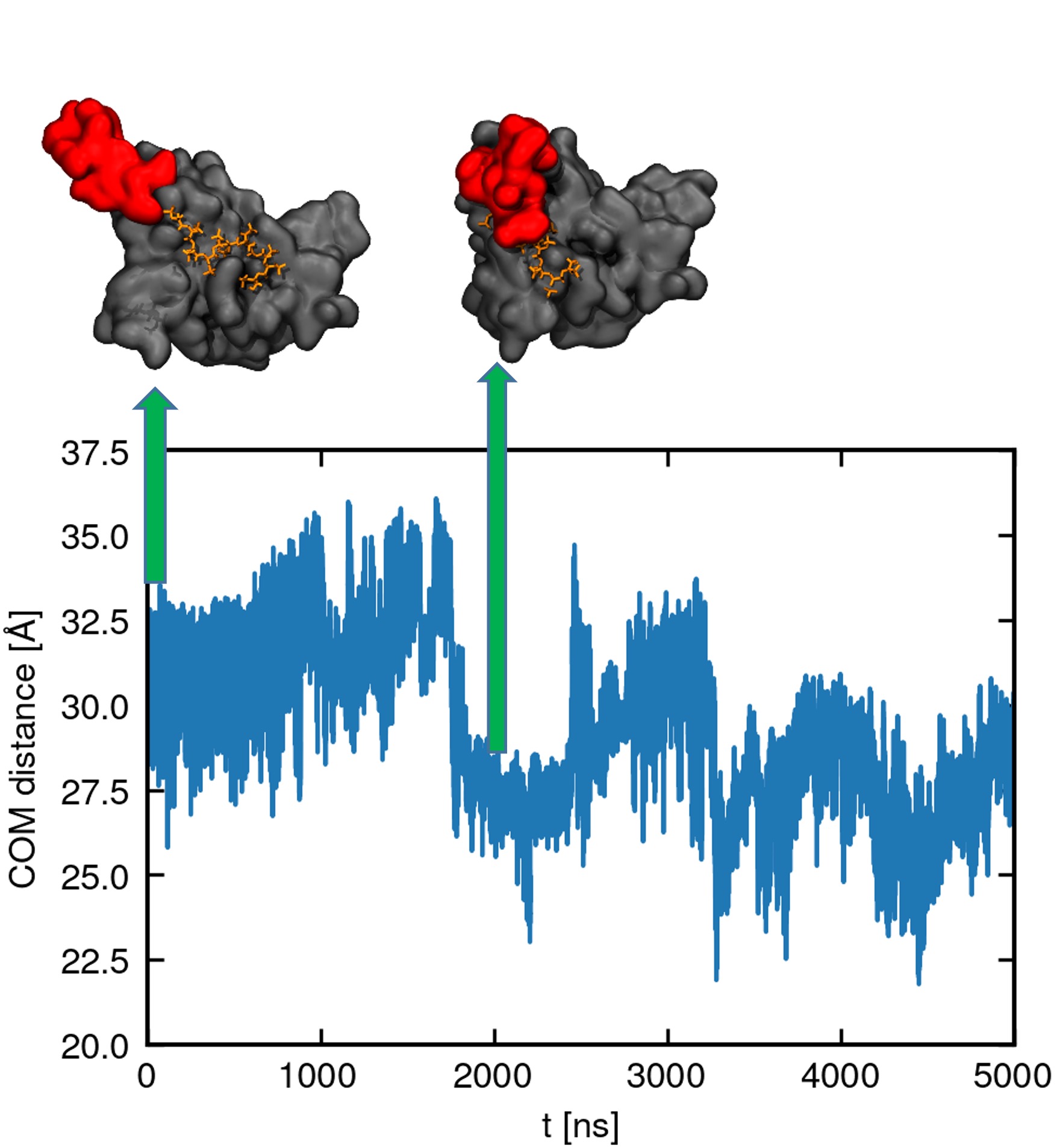}
    \caption{Time series of the distance between the center-of-mass (COM) of the loop region (residues 470 to 490) and the rest of the RBD from the simulation of the Delta variant and LPGS at a pulled distance $\xi =$ \SI{0.4}{\nano\meter} of the LPGS terminus from the RBD surface. Snapshots at \SI{0}{\nano\second} and \SI{2000}{\nano\second} are displayed above. LPGS is shown in orange and the RBD is shown in grey except the loop region in red. At around \SI{2000}{\nano\second}, LPGS is found to be entangled between the loop region and the rest part of the RBD.} 
    \label{fig:entanglement}
\end{figure}

\begin{figure}[H]
    \centering
    \includegraphics[width=0.62\textwidth]{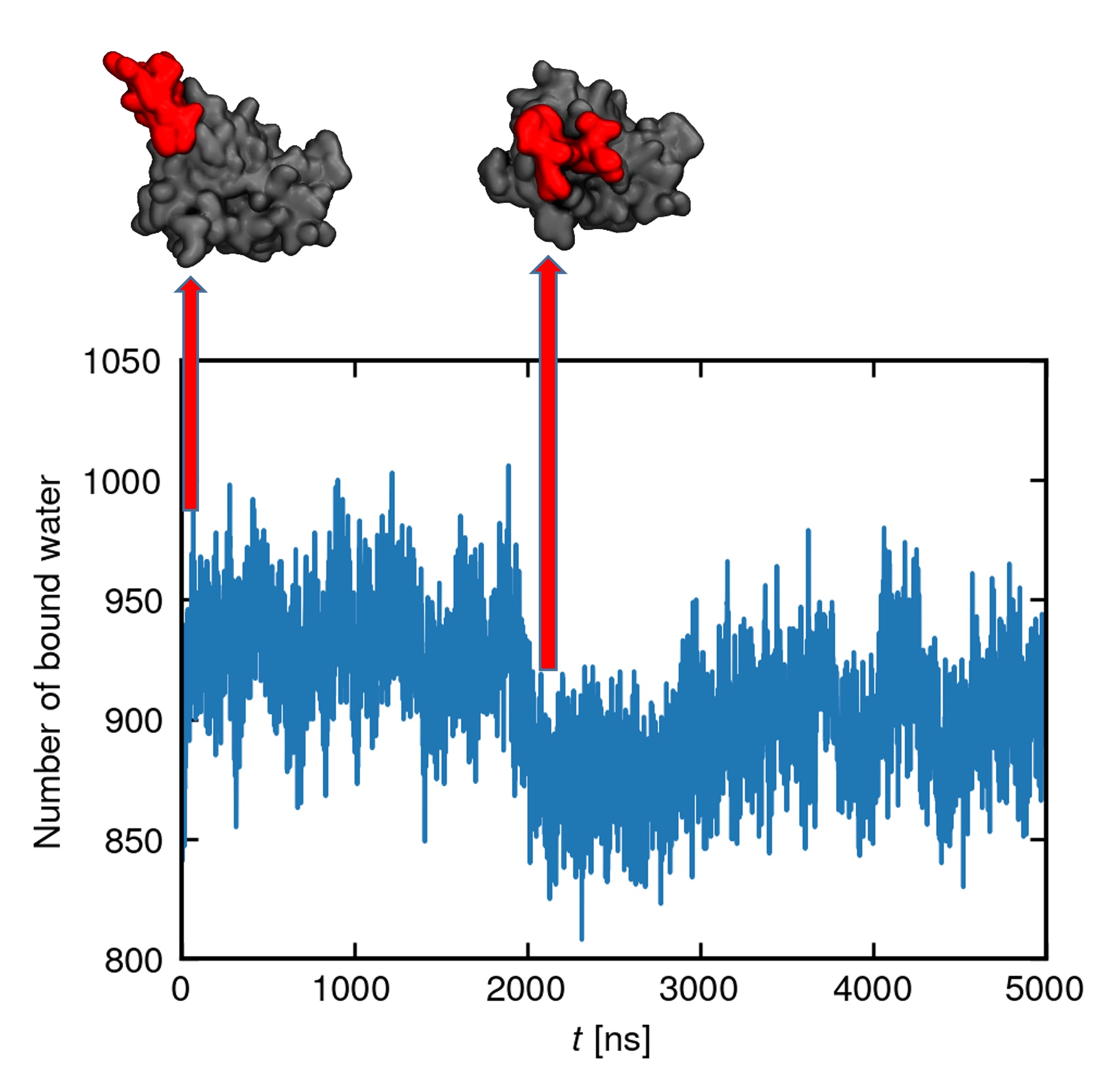}
    \caption{Time series of the number of bound water molecules to protein and LPGS from the simulation of the wild-type RBD at a pulled distance $\xi =$ \SI{2.8}{\nano\meter} of the LPGS terminus from the RBD surface. Snapshots at \SI{0}{\nano\second} and \SI{2000}{\nano\second} are displayed above the graph. The flexible loop region (residues 470 to 490) of the RBD is shown in red, whereas the rest part is in grey. LPGS is not shown for clarity.} 
    \label{fig:water_release_loop}
\end{figure}

\begin{figure}[H]
    \centering
    \includegraphics[width=1\textwidth]{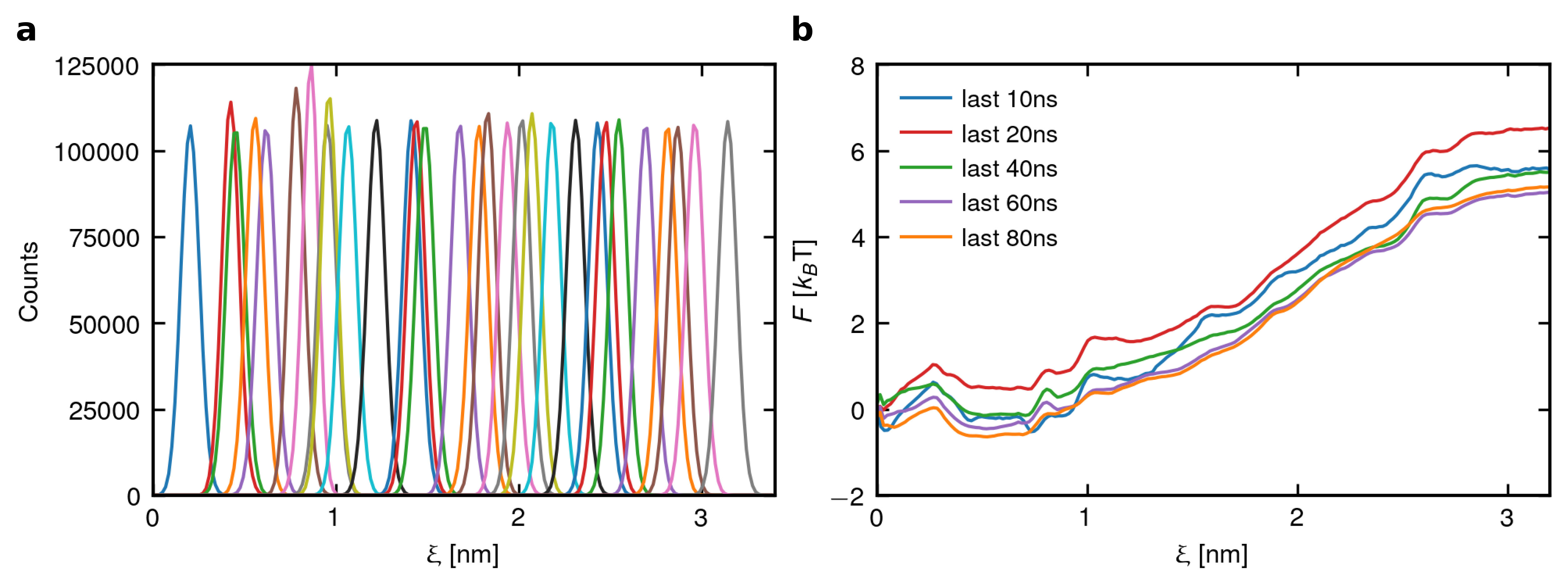}
    \caption{Convergence check for the umbrella sampling simulations for LPGS interactions with the wild-type RBD. \textbf{(a)} Histograms of different sampling windows showing nice overlap between consecutive windows. \textbf{(b)} Free energy profiles obtained using data from different time intervals of the simulations. The profiles are converged beyond 60 ns of simulation at each window.} 
    \label{fig:umbrella_blocks}
\end{figure}
\begin{figure}[H]
    \centering
    \includegraphics[width=\textwidth]{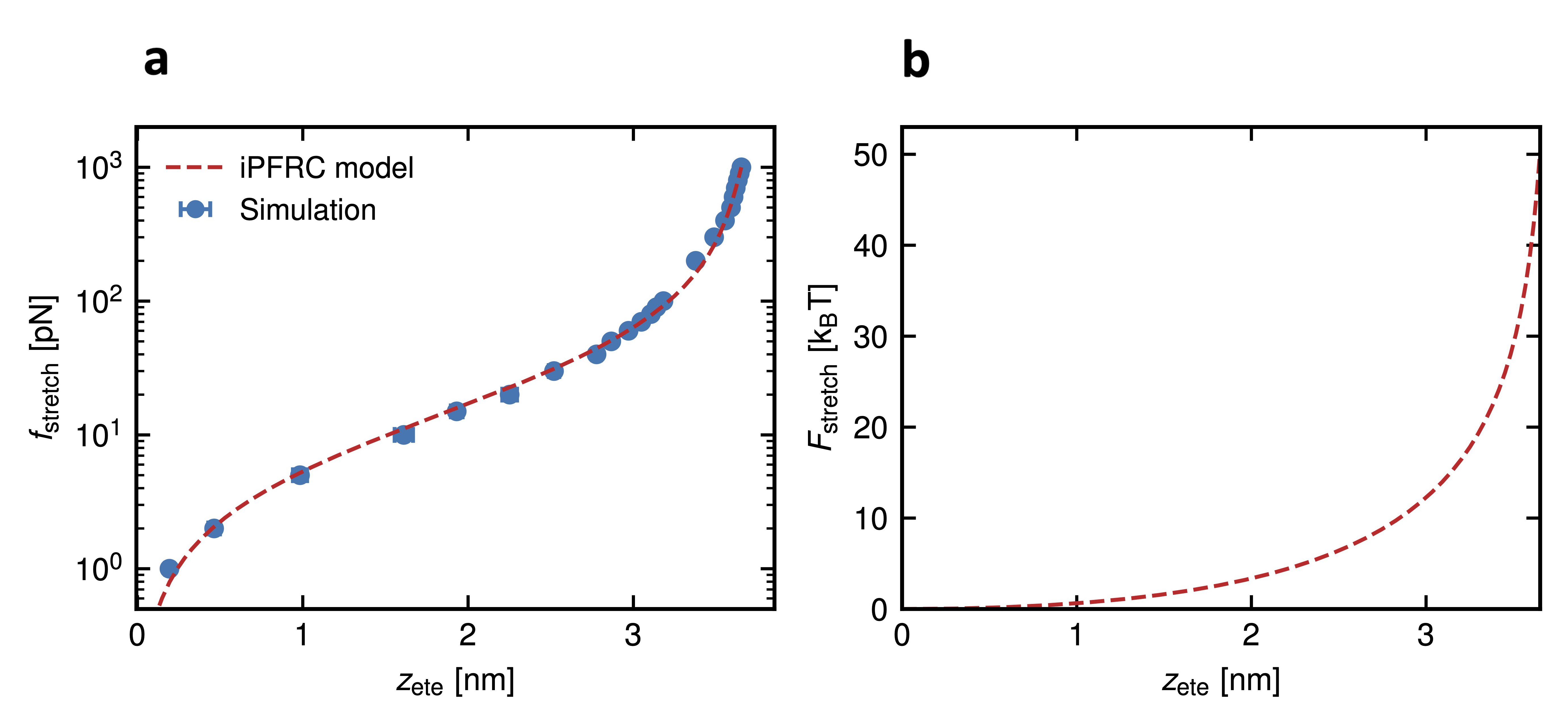}
    \caption{Stretching force and free-energy profiles for an LPGS undecamer. \textbf{(a)} Force--extension profile from the stretching simulations fitted with the iPFRC model. Not all error bars are visible since they are smaller than the symbol size. \textbf{(b)} Stretching free energy obtained by integrating the fitted iPFRC force--extension relation.}
    \label{fig:full_f-e_U_str}
\end{figure}
%
